\documentclass[reqno,12pt]{article}
\usepackage[english]{babel}

\usepackage{cite}
\usepackage{physics}
\usepackage{float}

\usepackage{ae} 
\usepackage[T1]{fontenc}
\usepackage{amsmath}
\usepackage{amssymb}
\usepackage{mathrsfs}
\usepackage{graphicx}
\usepackage{mdframed}
\usepackage{color}
\definecolor{darkblue}{cmyk}{0.9,0.9,0,0}
\usepackage[colorlinks=true,linkcolor=darkblue,citecolor=darkblue,urlcolor=darkblue]{hyperref}
\usepackage{epsfig}

\usepackage{graphicx}
\usepackage{tikz}

\newcommand{\beq}{\begin{equation}}
\newcommand{\eeq}{\end{equation}}
\newcommand\beqa{\begin{eqnarray}}
\newcommand\eeqa{\end{eqnarray}}
\newcommand\bea{\begin{array}}
\newcommand\eea{\end{array}}

\def\XXint#1#2#3{{\setbox0=\hbox{$#1{#2#3}{\int}$}
\vcenter{\hbox{$#2#3$}}\kern-.5\wd0}}

\newcommand{\neqa}{\nonumber\end{eqnarray}}

\newcommand{\<}{{\langle}}
\renewcommand{\>}{{\rangle}}

\newcommand{\re}{\relax{\rm I\kern-.18em R}}

\def\dd{\textrm{d}}

\def\su2{{SU(2)}}

\def\[{\left[}
\def\]{\right]}

\def\({\left(}
\def\){\right)}
\def\[{\left[}
\def\]{\right]}

\def\<{\langle}
\def\>{\rangle}

\def\i2{\frac{i}{2}}

	\def\i{\textrm{i}}

\usepackage{varioref}
\usepackage{makeidx}
\makeindex

\newcommand\blfootnote[1]{%
  \begingroup
  \renewcommand\thefootnote{}\footnote{\hspace{-6mm}#1}%
  \addtocounter{footnote}{-1}%
  \endgroup
}

\numberwithin{equation}{section}

\begin{document}

\thispagestyle{empty}

\renewcommand{\thefootnote}{\fnsymbol{footnote}}
\setcounter{page}{1}
\setcounter{footnote}{0}
\setcounter{figure}{0}

\vspace{-0.4in}

\begin{center}
$$$$
{\Large
Flat limit of massless scalar scattering in $\mathrm{AdS}_2$
\par}
\vspace{1.0cm}

{Sarthak Duary $^\text{\tiny 1}$}
\blfootnote{
\tt{sarthak.duary@gmail.com}}
\\ \vspace{1.2cm}
\footnotesize{
{\it $^\text{\tiny 1}$International Centre for Theoretical Sciences-TIFR,
	Shivakote, Hesaraghatta Hobli, Bengaluru North 560089, India  }\\
\vspace{4mm}
}
\end{center}


\begin{abstract}

We explore the flat limit of massless scalar scattering in $\mathrm{AdS}_2$. We derive the $1 \to 1$ $\mathcal{S}$-matrix from the CFT $2$-point function. We show a key property of the $2 \to 2$ $\mathcal{S}$-matrix in $2d$, where the contact interaction in the flat limit gives momentum conserving delta function. We show the factorization of the $n \to n$ $\mathcal{S}$-matrix for integrable models in the flat limit, focusing on contact interactions. We calculate the $\mathcal{S}$-matrix by linking the CFT operator on the AdS boundary to the scattering state in flat-space. We use bulk operator reconstruction to study massless scalar scattering in the flat limit and solve the Klein-Gordon equation in global $\mathrm{AdS}_2$ for the massless scalar field. The solution is simple, involving a pure phase in global time and a sinusoidal function in the radial coordinate. This simplicity also extends to the smearing function, allowing us to map the scattering state to the CFT operator while taking AdS corrections into account.\\

\end{abstract}

\newpage

\setcounter{page}{1}
\renewcommand{\thefootnote}{\arabic{footnote}}
\setcounter{footnote}{0}



{
\tableofcontents
}

\newpage

\section{Introduction}
\label{intro}

The theory of gravity, which encompasses various quantum fields and incorporates a negative cosmological constant, can be described as a weakly coupled QFT through the perturbative treatment of the curvature entwined with the AdS background. Despite the fact that the resulting QFT is AdS-based, we nevertheless continue to employ conventional methods of Witten diagram to calculate the AdS amplitudes using the the propagators (bulk-to-bulk propagator, and bulk-to-boundary propagator) and interaction vertices in AdS amplitudes. These AdS amplitudes are equivalent to the large-$N$ CFT correlation functions at the boundary of AdS, in accordance with the AdS/CFT correspondence \cite{Maldacena:1997re,Witten:1998qj,Gubser:1998bc}. Now, in our foray into implementing QFT on AdS space, we find that the incorporation of a large AdS radius, denoted as $L \to \infty$, becomes feasible by studying the effective Lagrangian. 

At first glance, by considering the effective Lagrangian, we might assume that taking the limit of a large AdS radius ($L \rightarrow \infty$) renders QFTs on AdS indistinguishable from those in flat space. We can also readily observe that as $L \rightarrow \infty$, the AdS background simplifies to flat space. 
This large AdS radius limit ($L \rightarrow \infty$) is aptly coined the flat limit. 

To be more concrete, we can map the CFT correlation function in $d$-dimensions into AdS amplitude in $(d+1)$-dimensions, and then we can take large AdS radius limit i.e., flat limit and we can relate the CFT correlation function with the $\mathrm{Mink}_{d+1}$ scattering amplitude:
\begin{equation}
	\mathrm{CFT_d} \leftrightarrow \mathrm{AdS_{d+1}} \xrightarrow{\text{Large AdS radius }} \mathrm{Mink_{d+1}}. \nonumber 
\end{equation}
It is difficult to incorporate AdS amplitudes into the aforementioned flat limit. The notion of the flat limit has been around for a while in the literature \cite{Polchinski:1999ry,Giddings:1999jq,Gary:2009ae,Gary:2009mi,Heemskerk:2009pn,Fitzpatrick:2010zm,Fitzpatrick:2011jn}. In \cite{Okuda:2010ym,Penedones:2010ue,Fitzpatrick:2011hu,Maldacena:2015iua, Hijano:2020szl, Komatsu:2020sag, Li:2021snj, Paulos:2016fap}, various flat limit mapping prescriptions were undertaken that were more detailed.  Utilizing AdS/CFT, when we apply the flat limit to AdS, it suggests that the correlation functions of the boundary CFT hold important information regarding the local bulk observable, $\mathcal{S}$-matrix of the corresponding flat-space theory. Nevertheless, there are plenty of schemes that have been developed supporting various CFT representations: (a) momentum space \cite{Raju:2012zr, Gadde:2022ghy}, (b) mellin space \cite{Penedones:2010ue,Fitzpatrick:2011hu,Paulos:2016fap, Li:2023azu}, and (c) coordinate space \cite{Okuda:2010ym,Maldacena:2015iua,Komatsu:2020sag, Hijano:2020szl, Li:2021snj, Paulos:2016fap, Duary:2022pyv, Duary:2022afn, Li:2022tby, Hijano:2019qmi}. 

In the case of a QFT residing in $(d+1)$-dimensional AdS spacetime while maintaining its isometries, the theory is governed by the $SO(d,2)$ conformal group. Consequently, we have the ability to define boundary correlation functions by considering correlation functions of local bulk operators. These operators are positioned closer to the conformal boundary through a process of maneuvering their insertion points. The complete collection of correlation functions establishes a theory that exhibits conformal invariance and exists solely on the boundary. When we place a local operator at the boundary of AdS, it acts as a source and gives rise to a particle inside. However, in the realm of the flat limit, this operator is mapped to the corresponding asymptotic state. 

In the flat limit, the central region of AdS space simply morphs into a flat-space. As a result, it becomes natural to envision the extraction of the $\mathcal{S}$-matrix of the flat-space QFT, from the correlation functions of the boundary CFT. The consequences arising from the flat limit manifest in striking dissimilarities between the phenomena of massless scattering and massive scattering. Massless particles are characterized by operators with a finite conformal dimension, whereas massive particles are characterized by operators with large conformal dimension $\Delta\sim L \rightarrow\infty$. This flat limit approach offers an alternative method to investigate the analytic structure of the flat-space $\mathcal{S}$-matrix, particularly in non-perturbative scenarios. It is advantageous because CFT possesses more constraints and their analytic structure is better comprehended, facilitating the analysis. 

\paragraph{Setup of the paper.}
The idea we persue in the paper is about exploring the flat limit of the AdS/CFT correspondence. Let us contemplate a scenario where, inside the $\mathrm{AdS}_2$ geometry, deep inside the center scattering phenomena unfurls. By confining our observations to scales diminutive in comparison to the characteristic length scale of $\mathrm{AdS}_2$, we can perceive $\mathrm{Mink}_2$ geometry. 

As a statement of the metric of the geometry, it simply means that the original $\mathrm{AdS}_2$ geometry, when we we examine the vicinity of the center of it, it will turn into $\mathrm{Mink}_2$ spacetime. The key idea here is that flat-space is a component of AdS space, implying that the physics must be encoded into AdS spacetime. Now, considering that the physics inside AdS spacetime is dual to CFT, it is reasonable to assert that CFT encapsulates the physics of flat-space in an additional dimension. This reasoning forms the fundamental logic behind the flat limit of AdS/CFT. Now, by taking the flat limit of the CFT correlation function, we confine the dual description of the correlation function, represented by bulk Witten diagrams, to a specific small region within AdS. In this region, the physics resembles that of flat-space, leading to the emergence of the flat-space $\mathcal{S}$-matrix from the CFT correlation function, as anticipated. 

\paragraph{Motivating questions of the paper.} 
\begin{itemize}
	\item One motivating question behind our work is to examine the kinematics that in $2d$, when two identical particles interact, their momenta remain unchanged between the initial and final states.
	\item Another motivation is to understand the delta functions, one for every set of incoming and outgoing momenta in the flat limit, which appear in higher-point scattering amplitudes in integrable models in $2d$. 
\end{itemize}

In this paper, we adhere to the underlying principles of the bulk operator reconstruction following \cite{Hamilton:2005ju,Hamilton:2006az} to study massless scalar scattering in the flat limit of $\mathrm{AdS_2}$, which presents a representation of the $\mathcal{S}$-matrix in flat-space.
For QFT in $\mathrm{AdS}_2$ see e.g.  \cite{Antunes:2021abs, Ouyang:2019xdd,Mazac:2016qev,Ferrero:2019luz,DiPietro:2017vsp,Maldacena:1998uz,Gross:2017vhb,Maldacena:2016hyu}.
In the paper, the representation of the $\mathcal{S}$-matrix for massless scalar involves smearing with a scattering smearing function and connecting it to the correlation function of the boundary. In the flat limit, the observer in the CFT cannot make usage of the with large conformal dimension limit $\Delta \rightarrow\infty$ for massless scattering.\footnote{$\Delta/L \sim m: \text{conformal dimension}~\Delta$ is finite for massless particle at the large AdS radius limit (in the flat limit), conformal dimension scales linearly with $L$, i.e., $\Delta \sim L \to \infty$ for massive particle.} The physics of flat spacetime is achieved instead by using the bulk-point limit. For massless scattering in $\mathrm{AdS}_4$, if the incoming particles are positioned at Lorentzian time $-\pi/2$, and outgoing particles are positioned at Lorentzian time $\pi/2$, the particles are zoomed in locally on the lightlike seperated bulk-point inside AdS, so that locally the amplitude looks like a flat-space amplitude. The bulk $\mathcal{S}$-matrix results as the coefficient of this singularity, which occurs when the correlator reaches this limit and singularizes on the appropriate sheet \cite{Heemskerk:2009pn,Maldacena:2015iua}. 
In the paper, we consider QFT in AdS with no gravity.  The configuration in this paper is distinct from the typical AdS/CFT scenario. In the typical AdS/CFT scenario, it is well-established that the presence of a boundary stress-energy tensor corresponds to the dynamic behavior of the bulk metric. In this paper, our focus is limited to examining QFT within a fixed AdS background geometry, disregarding the dynamical aspects typically considered in the AdS/CFT correspondence. This means that, unlike the standard AdS/CFT correspondence where the presence of a boundary stress-energy tensor corresponds to the dynamics of the bulk metric, our specific scenario involves a fixed bulk metric, thereby eliminating the existence of a stress-energy tensor in the boundary spectrum. A boundary conformal theory (BCT) is the term used to describe a conformally invariant theory on the boundary that lacks a stress-energy tensor. It is important to differentiate a BCT from the boundary conformal field theory (BCFT) which includes the existence of the stress-energy tensor.

We refer to the theory as boundary conformal theory (BCT) because there is no boundary stress-energy tensor associated with the dynamical bulk metric. In contrast, when a boundary stress-energy tensor is present, it typically corresponds to a dynamical bulk metric. In this paper, we limit our discussion to studying quantum field theory in a fixed AdS background, focusing on a theory without gravity. An example of BCT is a long-range spin chain, particularly in the continuum limit.

\paragraph*{An overview of the key findings.}
We will now give an overview of the key findings, followed by an organization of the paper. 
\begin{enumerate}
	\item \textbf{Bulk operator reconstruction in global $\mathrm{AdS}_2$.}
	In this paper, we focus on the reconstruction of an operator within the scattering region of AdS. Specifically, we concentrate on the reconstruction of the bulk operator corresponding to the massless scalar field in $\mathrm{AdS}_2$, using a first principles approach. 
	We first solve the Klein-Gordon equation for the massless scalar field within the global $\mathrm{AdS}_2$ setting. Remarkably, the solution is elegantly simple, exhibiting a pure phase in global time combined with a sinusoidal function dependent on the radial coordinate. This simplicity extends to the smearing function, which facilitates the mapping between the bulk and boundary operator. By expanding the smearing function under the large $L$ limit, we construct the creation or annihilation modes in terms of the CFT operators.
	
	\item 
	\textbf{Scattering state in a Fock space in the flat limit.}
	The involves selecting a local operator situated deep within the bulk of AdS, extracting a creation or annihilation mode through a straightforward Fourier transform, and subsequently taking a large radius of AdS limit to transition to the flat limit.
	\item \textbf{1 $\to$ 1 $\mathcal{S}$-matrix from the flat limit of CFT 2-point function.}
	By employing the map between the CFT operator and the flat-space scattering state, we evaluate the $\mathcal{S}$-matrix from the CFT $2$-point function which is proportional to the momentum conserving delta function $\delta(p_2-p_1)$. 
	
	\item \textbf{Kinematics of $2 \to 2$ massless $\mathcal{S}$-matrix in $2d$ in the flat limit.}
	In the flat limit, we see that the product of two delta functions for $2 \to 2$ massless $\mathcal{S}$-matrix in $2d$, naturally occur for $\phi^4$ contact interaction. 
	
	\item \textbf{Factorization of the $n \to n$ $\mathcal{S}$-matrix from the flat limit.}
	We demonstrate the factorization of the $n \to n ~\mathcal{S}$-matrix from the flat limit by focusing primarily on contact interaction, specifically at the tree level in leading order in coupling. 
	
	Furthermore, after AdS corrections are taken into account, we notice that delta functions conserving momentum present in the $\mathcal{S}$-matrix, up to subleading order of AdS corrections. We provide a qualitative analysis and discussion of this topic in the concluding remarks section.
\end{enumerate}


\paragraph{Organization of the paper.}
This paper is organized in the following way. In \S \ref{flatlimit}, we review the flat limit, 
make notations and conventions for succeeding sections. In \S \ref{solve}, we work out the solution of Klein-Gordon equation for the massless scalar field in global $\mathrm{AdS}_2$. We also calculate the bulk operator reconstruction smearing function for the massless scalar field in \S \ref{sf}. Then in \S \ref{Mapping}, we construct a map between a creation or an annhilation operator in flat-space, and a CFT operator, which also accounts for subleading order AdS corrections.  In \S \ref{warmup}, we evaluate the 1 $\to$ 1 $\mathcal{S}$-matrix from the flat limit of CFT 2-point function. In \S \ref{contact}, we calculate the 2 $\to$ 2 $\mathcal{S}$-matrix in the flat limit from the 4-point contact Witten diagram, and recover the kinematics of $2 \to 2$ scattering of identical particles in $2d$. In \S \ref{comp}, we evaluate the Witten diagram in global $\mathrm{AdS}_2$ coordinates. In \S \ref{factorization}, we comment on the factorization of the $n \to n$ $\mathcal{S}$-matrix from the flat limit. Finally, in \S \ref{cr}, we summarize our results and suggest future plans.

\section{The flat limit: $\mathrm{Mink}_2$ tiling inside $\mathrm{AdS}_{2}$ spacetime}
\label{flatlimit}
In this section, we review the basics facts about the flat limit of AdS, fix notations and conventions.  
\subsection{Notations and Conventions}
The objective of this section is to fix the notations and conventions for succeeding sections. 
We consider Lorentzian $\mathrm{AdS}_{2}$ space with global coordinate $(\tau,\rho)$. The embedding space coordinates $(X_1,X_2,X_3)$ is related to global coordinates as
\begin{equation}
	\begin{split}
		&X_1=L\frac{\cos\tau}{\cos\rho},~~X_2=L\frac{\sin\tau}{\cos\rho},~~X_3=L\tan\rho,
	\end{split}
\end{equation}
where $-\pi/2 <\rho <\pi/2, -\infty <\tau<\infty$. We note that AdS in global coordinates contains closed timelike curves, which pose a problem for causality. The existence of a compact timelike dimension can lead to the presence of closed timelike curves in the AdS spacetime. To maintain causality, we unwrap the global time coordinate $\tau$, allowing it to span the entire real line, i.e., $-\infty <\tau<\infty$. As the global time coordinate is no longer periodic, the closed timelike curves are eliminated from the spacetime. To summerize, we unwarp the AdS by completely extending the global time coordinate, thereby reaching the so-called covering space of AdS. 

We obtain Lorentzian $\mathrm{AdS}_{2}$ space from 3-dimensional Minkowski space $\mathbb{M}^{3}$ in (2, 1) signature. 
The embeddding space coordinates $X^M$ obey where, $M=1, 2, 3$.
\begin{equation}
	\begin{split}
		X_MX^M&=-L^2\\	
		\implies -X_1^2-X_2^2+X_3^2&=-L^2.	
	\end{split}	
\end{equation}
The metric in the embedding space coordinate is given by
\begin{equation}
	\dd s^2=-\dd X_1^2-\dd X_2^2+\dd X_3^2,
\end{equation}
which can be represented in the global coordinate system as follows	
\begin{equation}
	\dd s^2=\frac{L^2}{\cos^2\rho}\left(-\dd\tau^2+\dd\rho^2\right).
\end{equation}
The boundary of global $\mathrm{AdS}_2$ is at $\rho = \pm \frac{\pi}{2}$. The coordinate parametrizing the boundary of AdS is the global time $\tau$. We have
\begin{equation}
	\dd s_{\partial \mathrm{AdS}}^2 = \dd \tau^2.
\end{equation}
We begin by defining a specific region within $\mathrm{AdS}_{2}$ which, when subjected to a flat limit, transforms into a flat spacetime. The limit is achieved by letting the $\mathrm{AdS}_{2}$ length scale approach infinity. To begin, we establish a scattering region within the bulk of $\mathrm{AdS}_{2}$, which is analogous to a flat spacetime. 
Let us explore the process of taking the flat limit for the coordinates. In our notation, flat-space is represented by
\begin{equation}
	ds^2=-\dd t^2+\dd r^2.
\end{equation}
Taking the flat limit for coordinates is a straightforward process. We can achieve this by employing the coordinate transformation
\begin{equation}
	L\tan\rho=r~~,~~\tau L=t\,,
\end{equation}	
sending $L \rightarrow\infty$.
\section{Solution of Klein-Gordon equation in $\mathrm{AdS}_2$}
\label{solve}
In this section, we solve the Klein-Gordon equation for the massless scalar field in global $\mathrm{AdS}_2$. 
From the global $\mathrm{AdS}_2$ metric 
\begin{equation}
	\dd s^2=\frac{L^2}{\cos^2\rho}\left(-\dd\tau^2+\dd\rho^2\right).
\end{equation}

We consider a massless scalar field $\mathrm{\Phi}(\rho,\tau)$, that satisfies Klein-Gordon equation as given by
\begin{equation}
	\Box\mathrm{\Phi}=0.
\end{equation}
The d'Alembertian operator $\Box\mathrm{\Phi}$, corresponding to the global $\mathrm{AdS}_2$ metric can be expressed as 
\begin{equation}
	\label{box}
	\begin{split}
		\Box\mathrm{\Phi}
		&=\frac{\cos^2\rho}{L^2}\left[-\partial_\tau^2\mathrm{\Phi}+\partial^2_\rho\mathrm{\Phi}\right].
	\end{split}
\end{equation}		
Using separation of variables and requiring an oscillatory phase behaviour with global time implies solutions of the kind
\begin{equation}
	\mathrm{\Phi}(\rho,\tau)=e^{-\imath\omega \tau}R(\rho).
\end{equation}
We can express the radial equation satisfied by $R(\rho)$ as
\begin{equation}
	R''(\rho)+\omega^2R(\rho)=0.
\end{equation}
The solution for $R(\rho)$ takes the form
\begin{equation}
	R_{\omega}(\rho)={\mathcal{C}_1}_\omega\cos(\omega\rho)+{\mathcal{C}_2}_\omega\sin(\omega\rho).
\end{equation}
At the $\mathrm{AdS}_2$ boundary $\rho=\pm \frac{\pi}{2}$, we have $$R_\omega\left(\rho=\pm \frac{\pi}{2}\right)=0$$ which implies 
\begin{equation}
	R_n(\rho)\sim\sin(\omega_n\rho)
\end{equation}
where 
\begin{equation}
	\omega_n=2n~~\textmd{where},~~n\in\mathbb{N}.
\end{equation}
The discrete spectrum is given by $\omega_n=2n~~\textmd{where},~~n\in\mathbb{N}$. The solution can be expressed as follows
\begin{equation}
	\mathrm{\Phi}_n(\rho,\tau)=\mathfrak{c}_ne^{-2n\imath \tau}\sin(2n\rho),
\end{equation}
where $\mathfrak{c}_n$ is some normalization constant. $\mathfrak{c}_n$ can be fixed using the inner product
\begin{equation}
	\begin{split}	
		&\braket{\mathrm{\Phi}_n}{\mathrm{\Phi}_{n'}}\\&=\imath\int_{-\frac{\pi}{2}}^{\frac{\pi}{2}}\dd\rho\sqrt{-g}g^{\tau\tau} \eval{\left[\mathrm{\Phi}^*_n(\rho,\tau)~\partial_{\tau}\mathrm{\Phi}_{n'}(\rho,\tau)-\mathrm{\Phi}_{n'}(\rho,\tau)~\partial_{\tau}\mathrm{\Phi}^*_{n}(\rho,\tau)\right]}_{\tau=\textmd{constant}}\\
		&=2\pi\delta_{n,n'}.
	\end{split}	
\end{equation} 
which gives
\begin{equation}
	\mathfrak{c}_n=\sqrt{\frac{2}{n}}.
\end{equation}
We save the details of the computation of the normalization constant $\mathfrak{c}_n$ in Appendix \ref{normcons}.
\subsection{Bulk operator reconstruction smearing function for the massless scalar field}	
\label{sf}	
In this section, we evaluate the bulk operator reconstruction smearing function for the massless scalar field.
The scalar field $\mathrm{\Phi}(\rho,x)$ in $\mathrm{AdS}_2$ can be expressed as
\begin{equation}
	\begin{split}
		\mathrm{\Phi}(\rho,\tau)&=\sum_{n\in\mathbb{N}}\mathfrak{c}_n\sin(2n\rho)\left[e^{-2n\imath\tau}a_n+e^{+2n\imath\tau}a^{\dagger}_n\right].
	\end{split}
\end{equation}
where $a_{n}$ is the annihilation operator, and $a^{\dagger}_{n}$ is the creation operator. \\ At the $\mathrm{AdS}_2$ boundary, we define the boundary operator which is the primary CFT operator $\mathcal{O}(\tau')$ following the BDHM relation as \cite{Banks:1998dd}
\begin{equation}
	\begin{split}	
		\label{co}
		\mathcal{O}(\tau')&\equiv\lim_{\rho'\to\frac{\pi}{2}}\frac{\mathrm{\Phi}(\rho',\tau')}{\cos(\rho')}\\&=-\sum_{n\in\mathbb{N}}2n\mathfrak{c}_n\cos(\pi n)\left[e^{-2n\imath\tau'}a_n+e^{+2n\imath\tau'}a^{\dagger}_n\right].
	\end{split}
\end{equation}
In eq.\eqref{co}, we have the conformal dimension of dual primary CFT operator corresponding to massless scalar bulk field i.e., $\Delta=1$ since
\begin{equation}
	\Delta(\Delta-1)=m^2L^2.
\end{equation}
We get annihilation boundary operator $\mathcal{O}_+(\tau')$ and creation boundary operator $\mathcal{O}_-(\tau')$ as
\begin{equation}
	\label{ops}
	\begin{split}
		&\mathcal{O}_+(\tau')=-\sum_{n\in\mathbb{N}}2n\mathfrak{c}_n\cos(\pi n)e^{-2n\imath\tau'}a_n\\&\mathcal{O}_-(\tau')=-\sum_{n\in\mathbb{N}}2n\mathfrak{c}_n\cos(\pi n)e^{+2n\imath\tau'}a^{\dagger}_n.
	\end{split}
\end{equation}
By applying the inverse operation to the aforementioned relationships described in eq.\eqref{ops}, we obtain the following
\begin{equation}
	\label{recons}
	\begin{split}
		&a_{n}=-\frac{1}{2n\mathfrak{c}_n\cos(\pi n)}\int_{-\pi}^{\pi}\dd\tau'\mathcal{O}_+(\tau')e^{+2n\imath\tau'}\\&a^{\dagger}_{n}=-\frac{1}{2n\mathfrak{c}_n\cos(\pi n)}\int_{-\pi}^{\pi}\dd\tau'\mathcal{O}_-(\tau')e^{-2n\imath\tau'}.
	\end{split}
\end{equation}
By substituting the expression of the annihilation and creation operators into the function $\mathrm{\Phi}(\rho,\tau)$, we obtain
\begin{equation}
	\mathrm{\Phi}(\rho,\tau)=\int_{-\pi}^{\pi}\dd\tau'\left[K_+(\rho,\tau;\tau')\mathcal{O}_+(\tau')+K_-(\rho,\tau;\tau')\mathcal{O}_-(\tau')\right],
\end{equation}
where, the expression for the  bulk operator reconstruction smearing function $K_\pm(\rho,\tau;\tau')$ can be stated as follows
\begin{equation}
	\begin{split}
		K_\pm(\rho,\tau;\tau')&=-\sum_{n\in\mathbb{N}}\frac{1}{2n\cos(\pi n)}\sin(2n\rho)e^{\mp2n\imath\left(\tau-\tau'\right)}\\&=\sum_{n\in\mathbb{N}}\frac{(-1)^{2n+1}}{2n\cos(\pi n)}\sin(2n\rho)e^{\mp2n\imath\left(\tau-\tau'\right)}.
	\end{split}
\end{equation}	
In eq.\eqref{recons}, we reconstruct the local bulk operator in terms the boundary CFT operator.
In next section, we will calculate the modes of the massless scaler field in terms of the CFT operators in the flat limit. 
\section{Creation and annihilation operators:  massless scalar field}	
\label{Mapping}
In this section, we extract the modes like the creation and annihilation operators. By first principle, we calculate the map between creation and annihilation operators in flat-space and CFT operators in the boundary of $\mathrm{AdS}_2$. Deep within the center of $\mathrm{AdS}_2$, there exists a scattering region. We think of the massless scalar field $\mathrm{\Phi}(\rho,\tau)$ in $\mathrm{AdS}_2$ as the local bulk operator in the scattering region, which we reconstruct in terms of the boundary CFT operator in \S \ref{sf}.
In the flat limit, the field $\mathrm{\Phi}(r,t)$ is a free field in $2d$ flat-space. In order to construct it, we can combine the creation and annihilation operators linearly 
\begin{equation}
	\mathrm{\Phi}(r,t)=\int\frac{\dd p}{\sqrt{2\pi}}\frac{1}{\sqrt{2\omega_{p}}}\left(a_{p}\,e^{+\imath p\cdot x}+a^{\dagger}_{p}e^{-\imath p\cdot x}\right).
\end{equation}
The annihilation and creation operators can be extracted from the field $\mathrm{\Phi}(r,t)$ through a Fourier transform as 	
\begin{equation}
	\label{ca}
	\begin{split}
		&a_{p}=+\frac{\imath}{\sqrt{2\omega_{p}}}\int\frac{\dd r}{\sqrt{2\pi}}\,e^{-\imath p\cdot x}\overset{\leftrightarrow}{\partial_t}\mathrm{\Phi}(r,t)\\&a^{\dagger}_{p}=-\frac{\imath}{\sqrt{2\omega_{p}}}\int\frac{\dd r}{\sqrt{2\pi}}\,e^{+\imath p\cdot x}\overset{\leftrightarrow}{\partial_t}\mathrm{\Phi}(r,t),
	\end{split}
\end{equation}	
where $\overset{\leftrightarrow}{\partial_t}\equiv\overset{\rightarrow}{\partial_t}-\overset{\leftarrow}{\partial_t}$ and $p\cdot x=pr-\omega_{p}t$.

To successfully reconstruct the operator in the boundary of $\mathrm{AdS}_2$, a crucial step involves determining the dynamics of the operator. This can be likened to selecting an appropriate asymptotic Hamiltonian in the framework of flat space. In our analysis, we focus on free fields, which corresponds to the utilization of a free asymptotic Hamiltonian. This assumption, known as asymptotic decoupling in $\mathcal{S}$-matrix theory, leads us to the notion of a Fock space. By considering these principles, we pave the way for a coherent and meaningful reconstruction of the operator within the AdS boundary.
\subsection{ Flat limit and finite $L$ improvements to subleading order}
\label{finiteL}	
To have a flat limit, we need 
\begin{equation}
	\omega_n=2n=kL
\end{equation}
which gives
\begin{equation}
	\sum_{n\in\mathbb{N}}\to\frac{L}{2}\int\dd k.
\end{equation}
We substitute the following relation as
\begin{equation}
	\frac{t}{L}=\tau~~,~~\frac{r}{L}=\tan(\rho),
\end{equation}
in the smearing function, and get 
\begin{equation}
	\begin{split}
		K_\pm(r,t;\tau')&=-\frac{L}{2}\int\dd k\frac{(-1)^{kL}}{kL\cos(\frac{\pi kL}{2})}\sin(kL\tan[-1](\frac{r}{L}))e^{\mp kL\imath\left(\frac{t}{L}-\tau'\right)}\\&=-\int\frac{\dd k}{2k}\frac{e^{\mp \imath kt}}{\cos(\frac{\pi kL}{2})}e^{\imath kL\left(\pi\pm\tau'\right)}\sin(kL\tan[-1](\frac{r}{L})).
	\end{split}
\end{equation}	
We expand $\sin\Big(kL\tan[-1](\frac{r}{L})\Big)$ at large $L$ as
\begin{equation}
	\sin(kL\tan[-1](\frac{r}{L}))=\sin(kr)-\frac{kr^3}{3L^2}\cos(kr)+\mathcal{O}\left(\frac{1}{L^4}\right).
\end{equation}
Consequently, the smearing function can be derived up to subleading order in $\mathcal{O}\Big(\frac{1}{L^2}\Big)$ 
\begin{equation}
	\label{subleading1}
	\begin{split}
		K_\pm(r,t;\tau')&=-\int\frac{\dd k}{2k}\frac{e^{\mp \imath kt}}{\cos(\frac{\pi kL}{2})}e^{\imath kL\left(\pi\pm\tau'\right)}\sin(kr)\\&+\frac{r^3}{6L^2}\int\dd k\frac{e^{\mp \imath kt}}{\cos(\frac{\pi kL}{2})}e^{\imath kL\left(\pi\pm\tau'\right)}\cos(kr)\\
		&+\mathcal{O}\left(\frac{1}{L^4}\right).
	\end{split}
\end{equation}
In order to derive the creation/annihilation operators presented in eq.\eqref{ca}, we begin by examining the quantity
\begin{equation}
	\label{intn}
	\begin{split}
		&\int\dd r\,e^{-\imath p\cdot x}\overset{\leftrightarrow}{\partial_t}K_+(r,t;\tau')\\&=\int\dd r e^{-\imath pr+\imath\omega_{p}t}\left[\partial_tK_+(r,t;\tau')-\imath\omega_{p}K_+(r,t;\tau')\right].
	\end{split}
\end{equation}
By substituting the expression of the subleading order smearing function from eq.\eqref{subleading1}, the integrand in eq.\eqref{intn} without the exponential term can be expressed as follows
\begin{equation}
	\label{int1}
	\begin{split}
		&\partial_tK_+(r,t;\tau')-\imath\omega_{p}K_+(r,t;\tau')\\&=\imath\int\frac{\dd k}{2k}\frac{e^{- \imath kt}}{\cos(\frac{\pi kL}{2})}e^{\imath kL\left(\pi+\tau'\right)}\left(k+\omega_p\right)\sin(kr)\\&-\frac{\imath r^3}{12L^2}\int\dd k\frac{e^{- \imath kt}}{\cos(\frac{\pi kL}{2})}e^{\imath kL\left(\pi+\tau'\right)}\left(k+\omega_p\right)\cos(kr)\\
		&+\mathcal{O}\left(\frac{1}{L^4}\right).
	\end{split}
\end{equation}
The annihilation and creation operators up to subleading order in $\mathcal{O}\Big(\frac{1}{L^2}\Big)$ can be expressed as 
\begin{equation}
	\label{mapnew}
	\begin{split}
		&a_p=\frac{\imath}{4}\sqrt{\frac{\pi}{\omega_p}}\frac{1}{\cos(\frac{\pi\omega_{p} L}{2})}\left[1-\frac{1}{\omega_p^2L^2}+\mathcal{O}\left(\frac{1}{L^4}\right)\right]\\
		&\times  \int_{-\pi}^{\pi}\dd\tau'\,e^{\imath \omega_pL(\pi+\tau')}\mathcal{O}_+(\tau')\\
		&a^{\dagger}_p=-\frac{\imath}{4}\sqrt{\frac{\pi}{\omega_p}}\frac{1}{\cos(\frac{\pi\omega_{p} L}{2})}\left[1-\frac{1}{\omega_p^2L^2}+\mathcal{O}\left(\frac{1}{L^4}\right)\right]\\
		&\times  \int_{-\pi}^{\pi}\dd\tau'\,e^{\imath \omega_pL(\pi-\tau')}\mathcal{O}_-(\tau').
	\end{split}
\end{equation}
The details of the computation of the creation and annihilation operators as a smearing over global time of the CFT operators is worked out in Appendix \ref{app1}. 

We have the expression for $\mathrm{\Phi}$ as a function of CFT operators 
\begin{equation}
	\mathrm{\Phi}(\rho,\tau)=\int_{-\pi}^{\pi}\dd\tau'\left[K_+(\rho,\tau;\tau')\mathcal{O}_+(\tau')+K_-(\rho,\tau;\tau')\mathcal{O}_-(\tau')\right].
\end{equation}
Therefore, we basically perform a Fourier transform of a CFT operator smeared over the boundary of $\mathrm{AdS}_2$ in extracting the annihilation or creation operator. The map in eq.\eqref{mapnew} is a map between a creation or an annhilation operator in flat-space, and a CFT operator.

\section{Warm up exercise:~1 $\to$ 1 $\mathcal{S}$-matrix from the flat limit of CFT 2-point function}
\label{warmup}
\begin{figure}[H]
	\centering
	\includegraphics[width=0.8\linewidth]{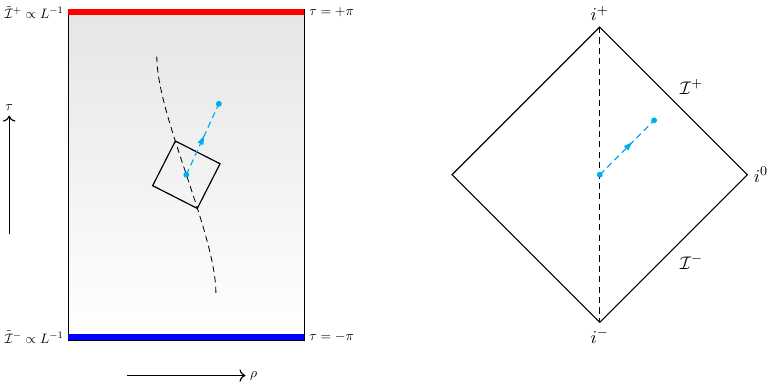}
	\caption{Dashed line represents the massless particle in the center of AdS. Massless scattering state corresponds to windows(red and blue regions in the CFT in the figure) of width $L^{-1}$, at $\tau=\pm \pi$. These red and blue portions of the CFT depict flat-space null infinity; it is where massless particles go and reach the AdS boundary.}
	\label{fig:Fig_1}
\end{figure}
In this section, we calculate the 1 $\to$ 1 $\mathcal{S}$-matrix from the flat limit of CFT 2-point function. The key component in the formulation of the $\mathcal{S}$-matrix is the scattering states. To make a transition from the AdS picture to the flat-space picture, we need to understand the relationship between operators and scattering states, specifically how we can construct the scattering states by smearing CFT operators. We establish the mapping of scattering states and CFT operators in the earlier section \ref{Mapping}. 

When taking the AdS radius to infinity to approach the flat-space limit, the exponential term in eq.\eqref{mapnew} becomes highly oscillatory (with a large phase). Consequently, the integral will be dominated by values of global time near $\pm \pi$. Essentially, this means that the integral is dominated by time intervals of $\mathcal{O}(1/L)$, where the global time is close to $\pm \pi$ within a window of size $1/L$. Thus, these regions of the AdS boundary function as null infinity.

In fig.\ref{fig:Fig_1}, the delineated red and blue regions indicate the specific locations where the insertion of a CFT operator is necessary. The red region corresponds to the insertion of a CFT operator associated with an outgoing massless scattering state, while the blue region represents the insertion of a CFT operator linked to an incoming massless scattering state. These strategic insertions play a crucial role in understanding the scattering dynamics within the depicted scenario.

We now can compute 1 $\to$ 1 $\mathcal{S}$-matrix from CFT $2$-point function. The formula expressing the mapping between the annihilation and creation modes and CFT operators derived in the earlier section \ref{Mapping} is given by
\begin{equation}
	\label{map}
	\begin{split}
		&a_p=\frac{\imath}{4}\sqrt{\frac{\pi}{\omega_p}}\frac{1}{\cos(\frac{\pi\omega_{p} L}{2})}\left[1-\frac{1}{\omega_p^2L^2}+\mathcal{O}\left(\frac{1}{L^4}\right)\right]\\& \times \int_{-\pi}^{\pi}\dd\tau'\,e^{\imath \omega_pL(\pi+\tau')}\mathcal{O}_+(\tau')\\
		&a^{\dagger}_p=-\frac{\imath}{4}\sqrt{\frac{\pi}{\omega_p}}\frac{1}{\cos(\frac{\pi\omega_{p} L}{2})}\left[1-\frac{1}{\omega_p^2L^2}+\mathcal{O}\left(\frac{1}{L^4}\right)\right]\\&\times  \int_{-\pi}^{\pi}\dd\tau'\,e^{\imath \omega_pL(\pi-\tau')}\mathcal{O}_-(\tau').
	\end{split}
\end{equation}
In this formula of eq.\eqref{map}, we take the AdS correction to the flat space creation and annihilation modes to subleading order, i.e., $\mathcal{O}\left(\frac{1}{L^2}\right)$. The global time integral is outshined by the operator insertions points $\tau^{\prime}=\pm \pi$ due to the extremely oscillatory nature of the exponential at the large AdS radius. 

Now, we construct ``in'' and ``out'' Fock space scattering states by acting the modes to the vacuum $\ket{0}$
\begin{equation}
	\begin{split}
		\ket{p, \text{in}}&= ~a^{\dagger}_{\text{in},p} \ket{0} \\
		\bra{\text{out},p}&= ~\bra{0}a_{\text{out},p}.
	\end{split}
\end{equation}
For 1 $\to$ 1 scattering, the $\mathcal{S}$-matrix which is the overlap between scattering states, in terms of the CFT $2$-point function is given by 
\begin{equation}
	\label{ft}
	\begin{split}
		\mathcal{S}_{1\to 1}&=\int\dd\tau_2'\,e^{\imath \omega_{p_2}L(\pi-\tau_2')}\int\dd\tau_1'\,e^{\imath \omega_{p_1}L(\pi+\tau_1')}\\&\langle\mathcal{O}_+(\tau_2')\mathcal{O}_-(\tau_1')\rangle \\
		&\times \frac{\pi   }{16 \sqrt{ \omega_{p_1} \omega_{p_2}}}\left[1-\frac{1}{L^2}\left( \frac{1}{\omega_{p_1}^2}+ \frac{1}{\omega_{p_2}^2}\right)+\mathcal{O}\left(\frac{1}{L^4}\right)\right]\\& \times  \sec(\frac{\pi  L \omega_{p_1}}{2}) \sec(\frac{\pi  L \omega_{p_2}}{2}).
	\end{split}
\end{equation}
Using the map between the CFT operator, and the scattering states in eq.\eqref{map}, we derive the map between  1 $\to$ 1 $\mathcal{S}$-matrix and CFT $2$-point function in eq.\eqref{ft}. 
Now, the Lorentzian CFT $2$-point function for the primary field is given by \cite{Fitzpatrick:2011jn} 
\begin{equation}
	\begin{split}
		\langle\mathcal{O}_+(\tau_2')\mathcal{O}_-(\tau_1')\rangle&=\frac{e^{\imath(\tau_2'-\tau_1')\Delta}}{\Big[1-e^{\imath(\tau_2'-\tau_1')}\Big]^{2\Delta}}\\
		&=\sum_{n\in\mathbb{N}}e^{-\imath(\tau_1'-\tau_2')(\Delta+n)}\frac{\mathrm{\Gamma}(2\Delta+n)}{\mathrm{\Gamma}(2\Delta)n!}.
	\end{split}
\end{equation}
We consider the integral
\begin{equation}
	\begin{split}
		&\int_{-\pi}^{\pi}\dd\tau_2'\,e^{\imath \omega_{p_2}L(\pi-\tau_2')}\int_{-\pi}^{\pi}\dd\tau_1'\,e^{\imath \omega_{p_1}L(\pi+\tau_1')}\,e^{-\imath(\tau_1'-\tau_2')(\Delta+n)}.
	\end{split}
\end{equation}
We change variables of integration as 
\begin{equation}
	\tau'_1=-\pi+\tilde{\tau}_1~~\textmd{and}~~\tau'_2=\pi-\tilde{\tau}_2.
\end{equation}
So, we have
\begin{equation}
	\begin{split}
		&\int_{0}^{2\pi}\dd\tilde{\tau}_2 e^{\imath\tilde{\tau}_2\left(\omega_{p_2}L-\left(\Delta+\frac{\omega_n}{2}\right)\right)}\int_{0}^{2\pi}\dd\tilde{\tau}_1 e^{\imath\tilde{\tau}_1\left(\omega_{p_1}L-\left(\Delta+\frac{\omega_n}{2}\right)\right)}e^{2\pi\imath \Delta}\\&=(2\pi)^2\delta_K\left(\omega_{p_2}L-\Delta-\frac{\omega_n}{2}\right)\delta_K\left(\omega_{p_1}L-\Delta-\frac{\omega_n}{2}\right)e^{2\pi\imath \Delta}
	\end{split}
\end{equation}
where $\omega_n=2n$.\\
We substitute $\Delta+n=\eta L$ which implies that for large $L$, we have
\begin{equation}
	\sum_{n\in\mathbb{N}}\to L\int\dd\eta.
\end{equation}
In this limit, the kroneker delta becomes
\begin{equation}
	\begin{split}
		&\delta_K\left(\omega_{p_2}L-\Delta-\frac{\omega_n}{2}\right)\delta_K\left(\omega_{p_1}L-\Delta-\frac{\omega_n}{2}\right)\\&\to\delta\left(\omega_{p_2}-\eta\right)\delta\left(\omega_{p_1}-\eta\right).
	\end{split}	
\end{equation}
Now, we have
\begin{equation}
	\frac{\mathrm{\Gamma}(2\Delta+n)}{\mathrm{\Gamma}(2\Delta)n!}\to\frac{1}{\mathrm{\Gamma}(2\Delta)}\frac{\mathrm{\Gamma}\left(\eta L+\Delta\right)}{\mathrm{\Gamma}\left(\eta L+1-\Delta\right)}
\end{equation}
which reduces to the following expression for large $L$ as 
\begin{equation}
	\label{gammafunc}
	\frac{\mathrm{\Gamma}(2\Delta+n)}{\mathrm{\Gamma}(2\Delta)n!}\to\frac{1}{\mathrm{\Gamma}(2\Delta)}\left(\eta L\right)^{2\Delta-1}.
\end{equation}
In deriving eq.\eqref{gammafunc} we use the series expansion of $\frac{\Gamma(x+a)}{\Gamma(x+b)}$ around $x=\infty$ which gives 
\begin{equation}
	\frac{\Gamma(x+a)}{\Gamma(x+b)}=x^{a-b}\Bigg(1+\mathcal{O}\Big(\frac{1}{x}\Big)\Bigg).
\end{equation}	
In conclusion, we obtain the following expression 
\begin{equation}
	\begin{split}
		&\int\dd\tau_2'\,e^{\imath \omega_{p_2}L(\pi-\tau_2')}\int\dd\tau_1'\,e^{\imath \omega_{p_1}L(\pi+\tau_1')}~\langle\mathcal{O}_+(\tau_2')\mathcal{O}_-(\tau_1')\rangle \\&=(2\pi)^2\frac{e^{2\pi\imath \Delta}}{\mathrm{\Gamma}(2\Delta)}L\int_{\eta=0}^{\infty}\dd\eta\,\delta\left(\omega_{p_2}-\eta\right)\delta\left(\omega_{p_1}-\eta\right)\left(\eta L\right)^{2\Delta-1}\\&=(2\pi)^2\frac{e^{2\pi\imath \Delta}}{\mathrm{\Gamma}(2\Delta)}L^{2\Delta}\omega^{2\Delta-1}_{p_1}\delta\left(\omega_{p_2}-\omega_{p_1}\right).
	\end{split}
\end{equation}
For the on-shell massless particles $\omega_{p_i}=p_i$.
So, the $1\to 1$ $\mathcal{S}$-matrix becomes
\begin{equation}
	\begin{split}
		\mathcal{S}_{1\to 1}&=\frac{\pi^3e^{2\pi\imath \Delta}}{4\mathrm{\Gamma}(2\Delta)}L^{2\Delta}\omega^{2(\Delta-1)}\sec[2](\frac{\pi L\omega}{2})\delta(p_2-p_1)\\&\times \left[1-\frac{2}{\omega^2L^2}+\mathcal{O}\left(\frac{1}{L^4}\right)\right]
	\end{split}	
\end{equation}
where $\omega=\omega_{p_1}=\omega_{p_2}$.
Now, the conformal dimension of dual primary CFT operator corresponding to massless scalar bulk field is $\Delta=1$ since
\begin{equation}
	\Delta(\Delta-1)=m^2L^2.
\end{equation}
The $\mathcal{S}$-matrix reduces to 
\begin{equation}
	\mathcal{S}_{1\to 1}=\frac{\pi^3}{4}L^{2}\sec[2](\frac{\pi L\omega}{2})\delta(p_2-p_1)\left[1-\frac{2}{\omega^2L^2}+\mathcal{O}\left(\frac{1}{L^4}\right)\right].
\end{equation}
Finally, the $\mathcal{S}$-matrix derived from the CFT $2$-point function is proportional to the momentum-conserving delta function, denoted as $\delta(p_2-p_1)$.
\section{$\mathcal{S}$-matrix in the flat limit from the 4-point contact Witten diagram}
\label{contact}
In this section, we evaluate the $\mathcal{S}$-matrix in the flat limit from the 4-point contact Witten diagram. We examine 4-point contact Witten diagram, where the interaction term is given by 
\begin{equation}
	\mathcal{L}=\lambda \phi^4,
\end{equation}
where, $\lambda$ is the coupling constant. 
The process of calculating Witten diagrams is made easier by employing the embedding space framework. Lorentzian global $\mathrm{AdS}_2$ is given by
\begin{equation}
	\dd s^2=\frac{L^2}{\cos[2](\rho)}\left(-\dd\tau^2+\dd\rho^2\right).
\end{equation}
The global $\mathrm{AdS}_2$ is embedded in $3$-dimensional Minkowski spacetime. The embedding coordinate $X$ is such that
\begin{equation}
	X^2=-L^2.
\end{equation}
The conformal boundary of AdS coordinate or CFT embedding coordinate can be thought of as null ray 
\begin{equation}
	P^2=0~~~,~~~P \sim \lambda P~~~(\lambda \in \mathbb{R}).
\end{equation}
The parametrization of the embedding coordinates $X$ and $P$ is given by
\begin{equation}
	\begin{split}
		X&=\frac{L}{\cos \rho}(\cos\tau,-\imath\sin\tau,\sin\rho)\\
		P&=(\cos\tau,-\imath\sin\tau,1).
	\end{split}
\end{equation}
The key constituent to evaluate the $\mathrm{AdS}_2$ $4$-point amplitude for contact interaction is the bulk-to-boundary propagator in $\mathrm{AdS}_2$  which is given by 
\begin{equation}
	\label{rep}
	\begin{split}
		G_{b\partial}(X,P)&=\frac{\mathcal{C}_{\Delta}}{(-2P.X/L)^{\Delta}}\\
		&=\frac{\mathcal{C}_{\Delta}}{\Gamma(\Delta)}\int_{0}^{\infty}\dd \mathfrak{t}~\mathfrak{t}^{\Delta-1} e^{ \frac{2\mathfrak{t}P.X}{L}},
	\end{split}
\end{equation}
where, the normalization constant is given by \cite{Freedman:1998tz,Fitzpatrick:2011ia}
\begin{equation}
	\mathcal{C}_{\Delta}=\frac{\Gamma(\Delta)}{2\sqrt{\pi}\Gamma\Big(\Delta+\frac{1}{2}\Big)}.
\end{equation}
The $4$-point amplitude in $\mathrm{AdS}_2$ associated with the contact interaction is expressed as follows
\begin{equation}
	\label{witten}
	\mathcal{A}=\lambda~\int_{\text{AdS}} \dd^{2}X \prod_{i=1}^{4} G_{b\partial}(X,P_i),
\end{equation}
where $G_{b\partial}$ is the bulk-to-boundary propagator 
\begin{equation}
	G_{b\partial}(X,P_i)=\frac{\mathcal{C}_{\Delta_i}}{(-2P_i.X/L)^{\Delta_i}}.
\end{equation}
Using the representation in eq.\eqref{rep} in eq.\eqref{witten}, we have
\begin{equation}
	\begin{split}
		\mathcal{A}&=\lambda L^2\Bigg(\prod_{i=1}^{4} \mathcal{C}_{\Delta_i}\Bigg) D_{\Delta_1\Delta_2\Delta_3\Delta_4}(P_i),
	\end{split}
\end{equation}
where, the $D$-function is given by
\begin{equation}
	\begin{split}
		D_{\Delta_1\Delta_2\Delta_3\Delta_4}(P_i)&=\prod_{i=1}^{4}\frac{1}{\Gamma(\Delta_i)}\int_{0}^{\infty}\prod_{i=1}^{4}\dd\mathfrak{t}_i \mathfrak{t}_i^{\Delta_i-1}\\
		&\times \int_{\text{AdS}}\dd(X/L)e^{-\frac{2\Big(\sum_{i=1}^{4}\mathfrak{t}_iP_i\Big).X}{L}}.
	\end{split}
\end{equation}
By performing the integration over the bulk $X$ coordinate, the $4$-point amplitude is simplified to the following expression
\begin{equation}
	\begin{split}	
		\mathcal{A}&=\lambda L^{2}\pi \Gamma\Bigg(\frac{\sum_{i=1}^4 \Delta_i-1}{2}\Bigg) \prod_{i=1}^{4}\frac{\mathcal{C}_{\Delta_i}}{\Gamma(\Delta_i)}\\& \times \int_{0}^{\infty}\prod_{i=1}^{4}\dd \mathfrak{t}_i ~\mathfrak{t}_i^{\Delta_i-1}e^{-\sum_{1\leq i<j\leq 4}\mathfrak{t}_i\mathfrak{t}_jP_{ij}},
	\end{split}
\end{equation}
where,
\begin{equation}
	\mathcal{C}_{\Delta_i}=\frac{\Gamma(\Delta_i)}{2\sqrt{\pi}~\Gamma\Big(\Delta_i+\frac{1}{2}\Big)}.
\end{equation}
\subsection{Witten diagram computation in global $\mathrm{AdS}_2$ coordinates}
\label{comp}
In this section, we will express the the $4$-point amplitude in global $\mathrm{AdS}_2$ coordinates. 
The boundary coordinates of the CFT are defined by the parametrization of null rays, resulting in the following expression
\begin{equation}
	\label{null}
	P_{ij}=-2P_i.P_j~,~\text{since }~P_i^2=0~,~P_j^2=0.
\end{equation}
We choose global time coordinates to parametrize $P_i$, and $P_j$ given by 
\begin{equation}
	\label{para}
	\begin{split}
		P_i&=(\cos\tau_i,-\imath\sin\tau_i,1)\\
		P_j&=(\cos\tau_j,-\imath\sin\tau_j,1).
	\end{split}
\end{equation}
Substituting eq.\eqref{para} in eq.\eqref{null} we get 
\begin{equation}
	P_{ij}=2(\cos(\tau_i-\tau_j)-1).
\end{equation}
Now, we restrict to massless scalars which are dual to conformal scalars of dimension $\Delta_i=1$.
As a result, the $4$-point amplitude in $\mathrm{AdS}_2$ is reduced to
\begin{equation}
	\mathcal{A}=\frac{\lambda L^2}{2\pi^{\frac{5}{2}}}\int_{0}^{\infty}\prod_{i=1}^{4}\dd\mathfrak{t}_i e^{-2\sum_{1\leq i<j\leq 4}\mathfrak{t}_i\mathfrak{t}_j(\cos(\tau_i-\tau_j)-1)}.
\end{equation}


\subsection{ 2 $\to$ 2 $\mathcal{S}$-matrix from well-chosen kinematics}
\label{wc}
In this section, we evaluate the 2 $\to$ 2 $\mathcal{S}$-matrix from well-chosen kinematics in the flat limit from the $4$-point amplitude. By examining a $\phi^4$ contact interaction, we can explicitly illustrate how the delta functions, which signify the equality of each pair of incoming and outgoing momenta, emerge naturally.

We construct ``in'' and ``out'' scattering states by acting the modes to the vacuum $\ket{0}$
\begin{equation}
	\begin{split}
		\ket{p_1,p_2; \text{in}}&= ~a^{\dagger}_{\text{in},p_1}a^{\dagger}_{\text{in},p_2} \ket{0} \\
		\bra{\text{out};p_3,p_4}&= ~\bra{0}a_{\text{out},p_3}a_{\text{out},p_4}.
	\end{split}
\end{equation}
For 2 $\to$ 2 scattering, the $\mathcal{S}$-matrix is 
\begin{equation}
	\begin{split}
		\mathcal{S}_{2\to 2}&=\bra{\text{out};p_3,p_4}\ket{p_1, p_2;\text{in}}\\&=\bra{0}a_{\text{out},p_3}a_{\text{out},p_4}a^{\dagger}_{\text{in},p_1}a^{\dagger}_{\text{in},p_2} \ket{0}.
	\end{split}	
\end{equation} 
The $\mathcal{S}$-matrix in terms of the CFT $4$-point function is given by 
\begin{equation}	
	\begin{split}
		\mathcal{S}_{2\to 2}&=\int\dd\tau_3'\,e^{\imath \omega_{p_3}L(\pi-\tau_3')}\int\dd\tau_4'\,e^{\imath \omega_{p_4}L(\pi-\tau_4')}\int\dd\tau_1'\,e^{\imath \omega_{p_1}L(\pi+\tau_1')}\int\dd\tau_2'\,e^{\imath \omega_{p_2}L(\pi+\tau_2')}\\
		&~\times \langle\mathcal{O}_+(\tau_4')\mathcal{O}_+(\tau_3')\mathcal{O}_-(\tau_2')\mathcal{O}_-(\tau_1')\rangle \times \frac{\pi ^2  }{64 \sqrt{ \omega_{p_1} \omega_{p_2}\omega_{p_3}\omega_{p_4}}}\\&\times \left[1-\frac{1}{L^2}\left( \frac{1}{\omega_{p_1}^2}+ \frac{1}{\omega_{p_2}^2}+\frac{1}{\omega_{p_3}^2}+\frac{1}{\omega_{p_4}^2}\right)+\mathcal{O}\left(\frac{1}{L^4}\right)\right]\\&~~\times\sec(\frac{\pi  L \omega_{p_1}}{2}) \sec(\frac{\pi  L \omega_{p_2}}{2})\sec(\frac{\pi  L \omega_{p_3}}{2}) \sec(\frac{\pi  L \omega_{p_4}}{2}).
	\end{split}
\end{equation}
With well-chosen kinematics, for $2\to 2$ scattering, the $4$-point function (modulo some overall factor)  
is written as
\begin{equation}
	\label{4pt}
	\begin{split}	
		&\langle\mathcal{O}_+(\tau_4')\mathcal{O}_+(\tau_3')\mathcal{O}_-(\tau_2')\mathcal{O}_-(\tau_1')\rangle\\&=\int_{0}^{\infty}\prod_{i=1}^{4}\dd \mathfrak{t}_i\,e^{-2\mathfrak{t}_1\mathfrak{t}_3\left(\cos(\tau'_1-\tau'_3)-1\right)}e^{-2\mathfrak{t}_2\mathfrak{t}_4\left(\cos(\tau'_2-\tau'_4)-1\right)}.
	\end{split}	
\end{equation}
We choose the kinematics in such a way that $P_{ij}=2(\cos(\tau_i-\tau_j)-1)=0$ for other $4$ out of ${4 \choose 2}=6$ pairings, i.e., $\{i, j\}$ as $\{1, 2\}, \{1, 4\}, \{2, 3\}, \text{and}, \{3, 4\}$.

Now, using the Jacobi-Anger expansion, we have
\begin{equation}
	\label{ja}
	e^{-z\cos(\theta)}=\sum_{n\in\mathbb{Z}}I_n(z)e^{-\imath n\theta}.
\end{equation}
Here, $I_n(z)$ is the $n$-th modified Bessel function of the first kind. 
So, using eq.\eqref{ja} the integrands in eq.\eqref{4pt}  becomes 
\begin{equation}
	e^{-2\mathfrak{t}_1\mathfrak{t}_3\left(\cos(\tau'_1-\tau'_3)-1\right)}=e^{2\mathfrak{t}_1\mathfrak{t}_3}\sum_{m\in\mathbb{Z}}I_m\left(2\mathfrak{t}_1\mathfrak{t}_3\right)e^{-\imath m\left(\tau'_1-\tau'_3\right)}
\end{equation}
and similarly
\begin{equation}
	e^{-2\mathfrak{t}_2\mathfrak{t}_4\left(\cos(\tau'_2-\tau'_4)-1\right)}=e^{2\mathfrak{t}_2\mathfrak{t}_4}\sum_{n\in\mathbb{Z}}I_n\left(2\mathfrak{t}_2\mathfrak{t}_4\right)e^{-\imath n\left(\tau'_2-\tau'_4\right)}.
\end{equation}
The $4$-point amplitude associated with the contact interaction, as given in eq.\eqref{4pt}, can be expressed as
\begin{equation}
	\begin{split}	
		&\langle\mathcal{O}_+(\tau_4')\mathcal{O}_+(\tau_3')\mathcal{O}_-(\tau_2')\mathcal{O}_-(\tau_1')\rangle\\&=\sum_{m,n\in\mathbb{Z}}C_{m,n}e^{-\imath m\left(\tau'_1-\tau'_3\right)}e^{-\imath n\left(\tau'_2-\tau'_4\right)}
	\end{split}	
\end{equation}
where the coefficients $C_{m,n}$ are defined as
\begin{equation}
	C_{m,n}\equiv\int_{0}^{\infty}\prod_{i=1}^{4}\dd \mathfrak{t}_i\ e^{2\mathfrak{t}_1\mathfrak{t}_3}e^{2\mathfrak{t}_2\mathfrak{t}_4}I_m\left(2\mathfrak{t}_1\mathfrak{t}_3\right)I_n\left(2\mathfrak{t}_2\mathfrak{t}_4\right).
\end{equation}
We examine the integral involved in the calculation 
\begin{equation}
	\label{subs}
	\begin{split}
		&\int_{-\pi}^{\pi}\dd\tau_3'\,e^{\imath \omega_{p_3}L(\pi-\tau_3')}\int_{-\pi}^{\pi}\dd\tau_1'\,e^{\imath \omega_{p_1}L(\pi+\tau_1')}\,e^{-\imath m(\tau_1'-\tau_3')}.
	\end{split}
\end{equation}
We substitute 
\begin{equation}
	\tau'_1=-\pi+\tilde{\tau}_1~~\textmd{and}~~\tau'_3=\pi-\tilde{\tau}_3
\end{equation}
in eq.\eqref{subs}, and we have
\begin{equation}
	\begin{split}
		&\int_{0}^{2\pi}\dd\tilde{\tau}_3 e^{\imath\tilde{\tau}_3\left(\omega_{p_3}L-\frac{\omega_m}{2}\right)}\int_{0}^{2\pi}\dd\tilde{\tau}_1 e^{\imath\tilde{\tau}_1\left(\omega_{p_1}L-\frac{\omega_m}{2}\right)}\\&=(2\pi)^2\delta_K\left(\omega_{p_3}L-\frac{\omega_m}{2}\right)\delta_K\left(\omega_{p_1}L-\frac{\omega_m}{2}\right)
	\end{split}
\end{equation}
where $\omega_m=2m$.\\ Similarly, we have
\begin{equation}
	\begin{split}
		&\int_{-\pi}^{\pi}\dd\tau_4'\,e^{\imath \omega_{p_4}L(\pi-\tau_4')}\int_{-\pi}^{\pi}\dd\tau_2'\,e^{\imath \omega_{p_2}L(\pi+\tau_2')}\,e^{-\imath n(\tau_2'-\tau_4')}\\&=(2\pi)^2\delta_K\left(\omega_{p_4}L-\frac{\omega_n}{2}\right)\delta_K\left(\omega_{p_2}L-\frac{\omega_n}{2}\right).
	\end{split}
\end{equation}
We substitute $m=\eta_m L$ and $n=\eta_nL$. In the limit of large $L$, we get 
\begin{equation}
	\sum_{m,n\in\mathbb{Z}}\to L^2\int\dd\eta_m\dd \eta_n.
\end{equation}
In this large $L$ limit, the kroneker delta becomes
\begin{equation}
	\begin{split}
		&\delta_K\left(\omega_{p_3}L-\frac{\omega_m}{2}\right)\delta_K\left(\omega_{p_1}L-\frac{\omega_m}{2}\right)\\
		&\delta_K\left(\omega_{p_4}L-\frac{\omega_n}{2}\right)\delta_K\left(\omega_{p_2}L-\frac{\omega_n}{2}\right)\\&\to\delta\left(\omega_{p_3}-\eta_m\right)\delta\left(\omega_{p_1}-\eta_m\right)\delta\left(\omega_{p_4}-\eta_n\right)\delta\left(\omega_{p_2}-\eta_n\right).
	\end{split}
\end{equation}
Now, we have $C_{m,n}\to C(L,\eta_m,\eta_n)$ in the large $L$ limit. 

\begin{figure}[H]
	\centering
	\includegraphics[width=0.7\linewidth]{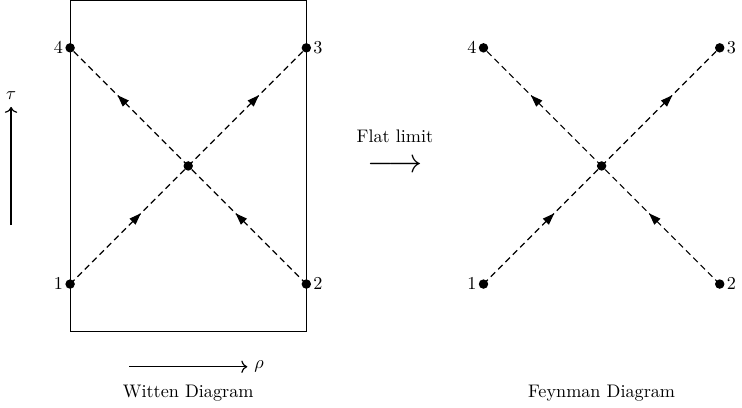}
	\caption{Witten diagram $\to$ Feynman diagram in the flat limit: contact diagram in $\mathrm{AdS}_2$ becomes $\delta\left(p_3-p_1\right)\delta\left(p_4-p_2\right)$ in the flat limit.}
	\label{fig:Fig_2}
\end{figure}

Hence, we obtain
\begin{equation}
	\begin{split}
		&\int\dd\tau_3'\,e^{\imath \omega_{p_3}L(\pi-\tau_3')}\int\dd\tau_4'\,e^{\imath \omega_{p_4}L(\pi-\tau_4')}\int\dd\tau_1'\,e^{\imath \omega_{p_1}L(\pi+\tau_1')}\\&\times \int\dd\tau_2'\,e^{\imath \omega_{p_2}L(\pi+\tau_2')}\times \langle\mathcal{O}_+(\tau_4')\mathcal{O}_+(\tau_3')\mathcal{O}_-(\tau_2')\mathcal{O}_-(\tau_1')\rangle\\&=(2\pi)^4L^2\int_{-\infty}^{\infty}\dd\eta_m\dd\eta_n\,C(L,\eta_m,\eta_n)\\
		&\times \delta\left(\omega_{p_3}-\eta_m\right)\delta\left(\omega_{p_1}-\eta_m\right)\delta\left(\omega_{p_4}-\eta_n\right)\delta\left(\omega_{p_2}-\eta_n\right)\\&=(2\pi)^4L^2C(L,\omega_{p_1},\omega_{p_2})\delta\left(\omega_{p_3}-\omega_{p_1}\right)\delta\left(\omega_{p_4}-\omega_{p_2}\right).
	\end{split}	
\end{equation}
For the on-shell massless particles $\omega_{p_i}=p_i$. So, the $2\to 2$ $S$-matrix is given by
\begin{equation}
	\label{flsm}
	\begin{split}
		\mathcal{S}_{2\to 2}&= \frac{\pi ^6  }{4  \omega_{p_1} \omega_{p_2}}L^2\sec[2](\frac{\pi  L \omega_{p_1}}{2}) \sec[2](\frac{\pi  L \omega_{p_2}}{2})\\&\times C(L,\omega_{p_1},\omega_{p_2})\delta\left(p_3-p_1\right)\delta\left(p_4-p_2\right)\\&\times\left[1-\frac{2}{L^2}\left( \frac{1}{\omega_{p_1}^2}+ \frac{1}{\omega_{p_2}^2}\right)+\mathcal{O}\left(\frac{1}{L^4}\right)\right].
	\end{split}
\end{equation}
The process of taking the flat limit effectively ``takes away the box'' from the contact Witten diagram in $\mathrm{AdS}_2$, turning it into a Feynman diagram characterized by the presence of delta functions $\delta\left(p_3-p_1\right)\delta\left(p_4-p_2\right)$ as in fig.\ref{fig:Fig_2}.	

Let's briefly spell out the kinematics of $2 \to 2$ scattering of identical particles in flat-space.
\paragraph{Kinematics of $2 \to 2$ scattering of identical particles in $2d$.} 
Let us consider a scenario where we begin with the simplest possible scattering process of $2 \to 2$ massless particles. Energy and momentum conservation determine 
\begin{equation}
	\begin{split}
		\{p_3, p_4\}&=\{p_1, p_2\}\\
		\{p_3, p_4\}&=\{p_2, p_1\}.
	\end{split}
\end{equation}
In other words, the set is the same, i.e., the set of initial momenta is equal to the set of the final momenta \cite{Paulos:2016but, Rosenhaus:2019utc}. When dealing with distinguishable particles, there exist two discernible alternatives. Nonetheless, when dealing with identical particles, they represent the same possibility. In the scenario of scattering between two indistinguishable particles in flat space, their momenta remain unaltered in both the initial and final states.
The $\mathcal{S}$-matrix becomes 
\begin{equation}
	\begin{split}
		\mathcal{S}_{2\to 2}& \propto \delta\left(p_3-p_1\right)\delta\left(p_4-p_2\right).
	\end{split}
\end{equation}
A noteworthy observation is that, due to the preservation of momentum in the scattering process, there is no significant distinction between the connected and disconnected components in the $\mathcal{S}$-matrix. This means that in $2d$, we can readily make transition between the connected and disconnected components without the need for complicated distributional considerations.

In the context of the flat limit, in eq.\eqref{flsm}, we demonstrate that the delta functions associated with the equality of each pair of incoming and outgoing momenta naturally emerge when considering a $\phi^4$ contact interaction. In eq.\eqref{flsm}, we also find that momentum-conserving delta functions show up in the $\mathcal{S}$-matrix up to subleading order, after accounting for AdS corrections. In the concluding remarks section, we provide a qualitative discussion concerning this issue.

\section{Comment on factorization of the $n \to n$ $\mathcal{S}$-matrix from the flat limit}
\label{factorization}
In this section, we will comment on the factorization of the $n \to n ~\mathcal{S}$-matrix from the flat limit.
As a result of the presence of \textit{higher conserved charges}, scattering amplitudes in integrable models, particularly higher-point $\mathcal{S}$-matrix, exhibit the inclusion of products of delta functions \cite{Zamolodchikov:1978xm, Dorey:1996gd, Parke:1980ki}. We denote the $n \to n$ $\mathcal{S}$-matrix by $\mathcal{S}_{n \to n}$, which can be expressed as
\begin{equation}
	\label{pair}
	\mathcal{S}_{n \to n} \propto \prod_{i=1}^{n} \delta(p^{\prime}_i-p_i)=\delta\left(p^{\prime}_1-p_1\right)\delta\left(p^{\prime}_2-p_2\right) \cdots  \left(p^{\prime}_n-p_{n}\right). 
\end{equation}
Here, in eq.\eqref{pair}, we denote $\prod_{i=1}^{2}\delta(p^{\prime}_i-p_i)$ which comes in the $2 \to 2$ $\mathcal{S}$-matrix. As in \S \ref{contact}, for $2 \to 2$ scattering, the $\mathcal{S}$-matrix is  
\begin{equation}
	\mathcal{S}_{2 \to 2}\propto \delta(p_3-p_1)\delta(p_4-p_2).
\end{equation}
Noting, $p_3=p^{\prime}_1, \text{and}~p_4=p^{\prime}_2$.
Now, we show that these delta functions appearing in eq.\eqref{pair} corresponding to equality of each pair of incoming and outgoing momenta naturally arise in the flat limit considering contact interaction $\phi^{2n}$. Although the factorization of the $\mathcal{S}$-matrix is non-perturbative statement, we show the factorization by only considering contact interaction i.e., to leading order in coupling (at tree level). At tree level, the contact interactions form the building blocks of the non-perturbative $\mathcal{S}$-matrix for the integrable models e.g., Sinh-Gordon model. The Lagrangian for the Sinh-Gordon model is as follows
\begin{equation}
	\mathcal{L}=\frac{1}{2}(\partial \phi)^2+\frac{m^2}{\mathfrak{b}^2}(\cosh (\mathfrak{b}\phi)-1).
\end{equation}
Through a perturbative expansion in the coupling parameter $b$ we get
\begin{equation}
	\mathcal{L}=\frac{1}{2}(\partial \phi)^2+\frac{1}{2} m^2\phi^2+\frac{m^2\mathfrak{b}^2}{4!}\phi^4+\frac{m^2\mathfrak{b}^4}{6!}\phi^6+\cdots.
\end{equation}


As described in section \ref{wc}, with well-chosen kinematics for $n\to n$ scattering, the $2n$-point function is written as 
\begin{equation}
	\begin{split}
		\label{2npt}
		\mathcal{G}_{2n}&=\int_{0}^{\infty}\prod_{i=1}^{n}\prod_{i^{\prime}=1}^{n}\dd \mathfrak{t}_i~\dd \mathfrak{t}_{i^{\prime}}e^{-2\mathfrak{t}_i\mathfrak{t}_{i^{\prime}}\left(\cos(\tau_{i}-\tau_{i^{\prime}})-1\right)}.
	\end{split}	
\end{equation}
We can do exactly same analysis to get the $n\to n ~\mathcal{S}$-matrix in the flat limit from CFT $2n$-point function which is given by
\begin{equation}
	\label{nton}
	\mathcal{S}_{n \to n} \propto \delta\left(p^{\prime}_1-p_1\right)\delta\left(p^{\prime}_2-p_2\right) \cdots  \left(p^{\prime}_n-p_{n}\right).
\end{equation}
The steps to get the resulting $n\to n ~\mathcal{S}$-matrix of eq.\eqref{nton} from CFT $2n$-point function are as follows.
\begin{itemize}
	\item First, we use Jacobi-Anger expansion in eq.\eqref{2npt} for each indivisual exponentials given by 
	\begin{equation}
		e^{-2\mathfrak{t}_i\mathfrak{t}_j\left(\cos(\tau_i-\tau_j)-1\right)}=e^{2\mathfrak{t}_i\mathfrak{t}_j}\sum_{\mathfrak{p}\in\mathbb{Z}}I_{\mathfrak{p}}\left(2\mathfrak{t}_i\mathfrak{t}_j\right)e^{-\imath \mathfrak{p}\left(\tau_i-\tau_j\right)}.
	\end{equation}
	\item We perform the global time integrals which gives kronecker deltas with sum. 
	\item In the large $L$ limit, the sum becomes integral, and each of the kronecker deltas becomes dirac delta like $\delta \left(\omega_{p_i}-\eta_{n_i}\right)$. The large $L$ limit plays the role of continuum limit which implies 
	\begin{equation}
		\sum_{n_i\in\mathbb{Z}}\to L\int\dd \eta_{n_i}.
	\end{equation}
	\item Finally, we use delta functions to reduce the $n\to n ~\mathcal{S}$-matrix in the flat limit yielding 
	\begin{equation}
		\mathcal{S}_{n \to n} \propto \delta\left(p^{\prime}_1-p_1\right)\delta\left(p^{\prime}_2-p_2\right) \cdots  \left(p^{\prime}_n-p_{n}\right).
	\end{equation}
\end{itemize}
In our analysis, we find that the momentum-conserving delta functions manifest in the $\mathcal{S}$-matrix up to subleading order, accounting for AdS corrections. We offer a qualitative discussion on this topic in the next concluding remarks section.
\section{Concluding remarks}
\label{cr}	
In this section, we wrap up by summarising our results and outlining potential avenues for future explorations. In this paper, we develop a mapping between scattering state and CFT operator that allows us to calculate the massless $\mathcal{S}$-matrix in $2d$ in the flat limit from the Lorentzian CFT $2$-point function. Our strategy of the mapping lies in the meticulous reconstruction of the bulk operator that corresponds to the massless scalar field in $\mathrm{AdS}_2$. Using the mapping between the CFT operator in the boundary of AdS and flat-space scattering state, we derive the 1 $\to$ 1 $\mathcal{S}$-matrix from the CFT $2$-point function, which is proportional to the momentum-conserving delta function $\delta(p_2-p_1)$. We demonstrate that the $2 \to 2$ massless $\mathcal{S}$-matrix in $2d$ exhibits a natural occurrence of the product of two delta functions due to the $\phi^4$ contact interaction in the flat limit. By concentrating mainly on contact interaction, specifically at the tree level, we show how the $n \to n ~\mathcal{S}$-matrix can be factorized from the flat limit. 

Additionally, even with AdS corrections, the momentum-conserving delta functions persist in the $\mathcal{S}$-matrix, albeit at subleading order.
Our prescription for calculating the $\mathcal{S}$-matrix in the flat limit and, AdS corrections to it is an alternative to the paper \cite{Gadde:2022ghy} exploring massive scalar particle scattering. They define the ``$\mathcal{S}$-matrix'' from the Witten diagram computation to only those terms which are proportianal to the overall momentum conserving delta function. We also retain the same term involving the momentum conserving delta functions to subleading order. The delta functions originating from the momentum conservation are a consequence of translational symmetry. While AdS lacks translational symmetry, the concept of momentum conservation can still be upheld within the flat-space region. The defining equation of the annihilation and creation operators in our approach serves as the rationale for retaining the momentum conserving delta functions. The defining equation of the annihilation and creation operators given by 	
\begin{equation}
	\label{def}
	\begin{split}
		&a_{p}=+\frac{\imath}{\sqrt{2\omega_{p}}}\int\frac{\dd r}{\sqrt{2\pi}}\,e^{-\imath p\cdot x}\overset{\leftrightarrow}{\partial_t}\mathrm{\Phi}(r,t)\\&a^{\dagger}_{p}=-\frac{\imath}{\sqrt{2\omega_{p}}}\int\frac{\dd r}{\sqrt{2\pi}}\,e^{+\imath p\cdot x}\overset{\leftrightarrow}{\partial_t}\mathrm{\Phi}(r,t).
	\end{split}
\end{equation}	
In inverting the creation and annihilation operators through this eq.\eqref{def}, we still mode expand the free field in terms of the decomposition of plane waves. In the context of full AdS spacetime, discussing the ``$\mathcal{S}$-matrix'' doesn't hold meaningful significance because the wave packet reaches the asymptotic boundary and returns in a finite time. Nonetheless, it is of significance to study the correlation function, which, given suitable kinematics, can be determined by the $\mathcal{S}$-matrix in the flat limit of spacetime. As the correlation function possesses a well-defined nature, we can study the AdS corrections, particularly focusing on subleading terms that relate to the correlation function within the flat spacetime framework.

\paragraph*{Factorization of the $n \to n$ $\mathcal{S}$-matrix. }
Concerning the aspect of factorization, we adhere to a less stringent interpretation of the $n \to n$ $\mathcal{S}$-matrix denoted as $\mathcal{S}_{n \to n}$. This can be represented by the expression
\begin{equation}
	\mathcal{S}_{n \to n} \propto \prod_{i=1}^{n} \delta(p^{\prime}_i-p_i)=\delta\left(p^{\prime}_1-p_1\right)\delta\left(p^{\prime}_2-p_2\right) \cdots  \left(p^{\prime}_n-p_{n}\right). 
\end{equation}

When it comes to the factorization of $2 \to 2$ scattering in $2d$ flat-space for processes like $n\to n$, relying solely on contact interactions described by $\phi^{2n}$ is usually insufficient. As demonstrated in a tree-level analysis in \cite{Dorey:1996gd}, both the $\phi^4$ and $\phi^6$ coefficients must be adjusted to achieve the cancellation of the connected $2 \to 4$ amplitude, equivalent to the complete factorization of $3 \to 3$. This adjustment is crucial to cancel out the connected $2 \to 4$ amplitude, a requirement equivalent to achieving the complete factorization of $3 \to 3$. This fine-tuning process leads to the emergence of potentials like $\cos (\beta \phi)$, as observed in models such as Sine-Gordon, ensuring quantum factorization and integrability.

Concerning contact interactions, it is crucial to observe that at the tree level, diagrams featuring two quartic vertices also contribute to the $3 \to 3$ amplitude. In maintaining a consistent loop expansion, one must account for a $\phi^6$ coupling alongside two powers of a $\phi^4$ coupling. The combination of these two diagram types contact, and exchange is what is analyzed in the tree-level integrability in \cite{Dorey:1996gd} (refer to pages 2 and 3 of \cite{Dorey:1996gd}).


Analyzing the tree level diagram for $\phi^4$ in $\mathrm{AdS}_2$ when it comes to the $3 \to 3$ process is more intricate compared to the contact $\phi^6$. Due to this, we consider only contact interactions in the paper. 
At the tree level, it is important to include both the exchange $\phi^4$ $3\to 3$ diagram and a contact diagram involving $\phi^6$ interaction for $3 \to 3$ process. In addition to decomposing the $\mathcal{S}$-matrix into a product of delta-functions, it is crucial to account for the coefficients originating from these diagram classes to demonstrate factorizability. Our analysis focuses on proving the product of delta-functions, where these delta-functions serve as an overall factor, and the contact diagram will not disrupt the overall product of delta-functions.


Finally, we move on to talk about some fascinating open problems. 
\subsection{Future plans}	
\paragraph{Integrable bootstrap from the flat limit of AdS/CFT.}	

One interesting potential application lies in the exploration of what insights the flat limit of AdS/CFT can provide regarding non-perturbative $\mathcal{S}$-matrices. In $2d$, there are plethora of exactly solvable $\mathcal{S}$-matrices in flat-space. If the exactly solvable or integrable QFTs are placed in $\mathrm{AdS}_2$, the interesting question arises: does integrability persist in some manner, meaning, do subleading corrections to the flat-space $\mathcal{S}$-matrix retain integrable properties?
Following a similar line of thought, it would be captivating to investigate those models which have exactly solvable $\mathcal{S}$-matrices in $\mathrm{AdS}_2$ in terms of boundary correlators. Different integrable models in $\mathrm{AdS_2}$ are studied in \cite{Giombi:2017cqn,Beccaria:2019dws,Beccaria:2019stp,Beccaria:2019ibr,Beccaria:2019mev,Beccaria:2019dju,Beccaria:2020qtk} but, these attempts are at the level of perturbative boundary correlators. It is intriguing to contemplate a notion of integrability for the non-pertubative boundary correlators.

It would be interesting to generalize the mapping between $\mathrm{CFT}_1$ correlators and flat-space $S$-matrices for massive scalar fields in $\mathrm{AdS}_2$, and make a connection with the recent developments of conformal bootstrap \cite{Cordova:2022pbl}. It would also be exciting to make a connection between Celestial $\mathrm{CFT}_0$ amplitudes developed in \cite{Duary:2022onm, Kapec:2022xjw} from the flat limit of $\mathrm{CFT}_1$ correlators.


\subsection*{Acknowledgements}

I thank  Nava Gaddam, R. Loganayagam, Pronobesh Maity, and definitely Pabitra Ray for discussions and/or related collaboration. I express my gratitude for the support received from the Department of Atomic Energy, Government of India, through project number RTI4001.

\appendix
\section{Calculation of the normalization constant of the massless scalar field in $\mathrm{AdS}_2$}
\label{normcons}
In this Appendix, we evaluate the normalization constant of the massless scalar field in $\mathrm{AdS}_2$. 
The solution for the massless scalar field in $\mathrm{AdS}_2$ is expressed as
\begin{equation}
	\mathrm{\Phi}_n(\rho,\tau)=\mathfrak{c}_ne^{-2n\imath \tau}\sin(2n\rho),
\end{equation}
where $\mathfrak{c}_n$ is the normalization constant. We evaluate $\mathfrak{c}_n$ using the inner product
\begin{equation}
	\begin{split}	
		\label{norm}
		&\braket{\mathrm{\Phi}_n}{\mathrm{\Phi}_{n'}}\\&=\imath\int_{-\frac{\pi}{2}}^{\frac{\pi}{2}}\dd\rho\sqrt{-g}g^{\tau\tau}\eval{\left[\mathrm{\Phi}^*_n(\rho,\tau)~\partial_{\tau}\mathrm{\Phi}_{n'}(\rho,\tau)-\mathrm{\Phi}_{n'}(\rho,\tau)~\partial_{\tau}\mathrm{\Phi}^*_{n}(\rho,\tau)\right]}_{\tau=\textmd{constant}}\\
		&=2\pi\delta_{n,n'}.
	\end{split}	
\end{equation} 
For the $\mathrm{AdS}_2$ metric the determinant is given by 
\begin{equation}
	\begin{split}
		g&=\Big(-\frac{L^2}{\cos^2\rho}\Big)\Big(\frac{L^2}{\cos^2\rho}\Big)=-L^4 \sec^4\rho\\
		\implies \sqrt{-g}&=L^2 \sec^2\rho.
	\end{split}
\end{equation}
and 
\begin{equation}
	g^{\tau \tau}=-\frac{1}{L^2\sec^2\rho}.
\end{equation}
From eq.\eqref{norm}, we have
\begin{equation}
	\label{det}
	\begin{split}
		&\imath\int_{-\frac{\pi}{2}}^{\frac{\pi}{2}}\dd\rho\sqrt{-g}g^{\tau\tau}\eval{\left[\mathrm{\Phi}^*_n(\rho,\tau)~\partial_{\tau}\mathrm{\Phi}_{n}(\rho,\tau)-\mathrm{\Phi}_{n}(\rho,\tau)~\partial_{\tau}\mathrm{\Phi}^*_{n}(\rho,\tau)\right]}_{\tau=\textmd{constant}}\\
		&=2\pi
	\end{split}
\end{equation}
From eq.\eqref{det}
\begin{equation}
	\mathfrak{c}_n=\sqrt{\frac{2}{n}}.
\end{equation}
We use the integral 
\begin{equation}
	\int_{-\frac{\pi}{2}}^{\frac{\pi}{2}}\sin^2 (2n\rho) \dd\rho=\frac{\pi}{2},
\end{equation}
where, $n\in\mathbb{N}.$
\section{Derivation of the formula for the creation and annihilation operators of massless scalar field in terms of the CFT operators}
\label{app1}
In this Appendix, we derive the formula for the creation and annihilation operators restricting to massless scalar field in $\mathrm{AdS}_2$ as a smearing over global time of the CFT operators against some exponential.
We note the smearing function to subleading order in $\mathcal{O}\Big(\frac{1}{L^2}\Big)$ from section \ref{finiteL}
\begin{equation}
	\label{subleading}
	\begin{split}
		K_\pm(r,t;\tau')&=-\int\frac{\dd k}{2k}\frac{e^{\mp \imath kt}}{\cos(\frac{\pi kL}{2})}e^{\imath kL\left(\pi\pm\tau'\right)}\sin(kr)\\&+\frac{r^3}{6L^2}\int\dd k\frac{e^{\mp \imath kt}}{\cos(\frac{\pi kL}{2})}e^{\imath kL\left(\pi\pm\tau'\right)}\cos(kr)\\
		&+\mathcal{O}\left(\frac{1}{L^4}\right).
	\end{split}
\end{equation}
In eq.\eqref{subleading}, we expand the smearing function to subleading order in $\mathcal{O}\Big(\frac{1}{L^2}\Big)$.	
The annihilation and creation operators in terms of the scalar fields are given by 	
\begin{equation}
	\label{ca1}
	\begin{split}
		&a_{p}=+\frac{\imath}{\sqrt{2\omega_{p}}}\int\frac{\dd r}{\sqrt{2\pi}}\,e^{-\imath p\cdot x}\overset{\leftrightarrow}{\partial_t}\mathrm{\Phi}(r,t)\\&a^{\dagger}_{p}=-\frac{\imath}{\sqrt{2\omega_{p}}}\int\frac{\dd r}{\sqrt{2\pi}}\,e^{+\imath p\cdot x}\overset{\leftrightarrow}{\partial_t}\mathrm{\Phi}(r,t),
	\end{split}
\end{equation}	
where $\overset{\leftrightarrow}{\partial_t}\equiv\overset{\rightarrow}{\partial_t}-\overset{\leftarrow}{\partial_t}$ and $p\cdot x=pr-\omega_{p}t$.
To obtain the creation/annhilation operators in eq.\eqref{ca1}, first we consider the quantity 
\begin{equation}
	\label{int}
	\begin{split}
		&\int\dd r\,e^{-\imath p\cdot x}\overset{\leftrightarrow}{\partial_t}K_+(r,t;\tau')\\&=\int\dd r e^{-\imath pr+\imath\omega_{p}t}\left[\partial_tK_+(r,t;\tau')-\imath\omega_{p}K_+(r,t;\tau')\right].
	\end{split}
\end{equation}
Now, after substituting the expression of the smearing function of eq.\eqref{subleading} to subleading order in $\mathcal{O}\Big(\frac{1}{L^2}\Big)$, the integrand without the exponential piece of eq.\eqref{int} becomes  
\begin{equation}
	\label{int1}
	\begin{split}
		&\partial_tK_+(r,t;\tau')-\imath\omega_{p}K_+(r,t;\tau')\\&=\imath\int\frac{\dd k}{2k}\frac{e^{- \imath kt}}{\cos(\frac{\pi kL}{2})}e^{\imath kL\left(\pi+\tau'\right)}\left(k+\omega_p\right)\sin(kr)\\
		&-\frac{\imath r^3}{12L^2}\int\dd k\frac{e^{- \imath kt}}{\cos(\frac{\pi kL}{2})}e^{\imath kL\left(\pi+\tau'\right)}\left(k+\omega_p\right)\cos(kr)\\
		&+\mathcal{O}\left(\frac{1}{L^4}\right).
	\end{split}
\end{equation}
Substituting eq.\eqref{int1}, the integral of eq.\eqref{int} reduces to the following integral 
\begin{equation}
	\label{piece1}
	\begin{split}
		\int\dd r\,e^{-\imath pr}\sin(kr)&=\int\dd r\,e^{-\imath \omega_{p}r}\sin(kr)=\frac{\pi}{2\imath}\delta(k-\omega_p)\\
		&~~~\textmd{for}~~k>0.
	\end{split}
\end{equation}
We then differentiate eq.\eqref{piece1} by $k$ three times, and we have
\begin{equation}
	\begin{split}	
		\int\dd r\,e^{-\imath pr}r^3\cos(kr)&=\int\dd r\,e^{-\imath \omega_{p}r}r^3\cos(kr)\\&=-\frac{\pi}{2\imath}\delta^{(3)}(k-\omega_p)\\&=\frac{3\pi}{\imath k^3}\delta(k-\omega_p)~~~~\textmd{for}~~k>0.
	\end{split}	
\end{equation}
Finally, eq.\eqref{int} gives 
\begin{equation}	
	\begin{split}
		&\int\dd r\,e^{-\imath p\cdot x}\overset{\leftrightarrow}{\partial_t}K_+(r,t;\tau')\\&=\frac{\pi}{2}\frac{e^{\imath \omega_pL(\pi+\tau')}}{\cos(\frac{\pi\omega_p L}{2})}-\frac{\pi}{2\omega_p^2L^2}\frac{e^{\imath \omega_pL(\pi+\tau')}}{\cos(\frac{\pi\omega_p L}{2})}+\mathcal{O}\left(\frac{1}{L^4}\right).
	\end{split}
\end{equation}
Therefore, we have the annihilation operator as 	
\begin{equation}
	\begin{split}	
		a_p&=\frac{\imath}{4}\sqrt{\frac{\pi}{\omega_p}}\frac{1}{\cos(\frac{\pi\omega_{p} L}{2})}\left[1-\frac{1}{\omega_p^2L^2}+\mathcal{O}\left(\frac{1}{L^4}\right)\right]\\&\times \int_{-\pi}^{\pi}\dd\tau'\,e^{\imath \omega_pL(\pi+\tau')}\mathcal{O}_+(\tau'),
	\end{split}
\end{equation}
and similarly we have the creation operator as 	
\begin{equation}
	\begin{split}		
		a^{\dagger}_p&=-\frac{\imath}{4}\sqrt{\frac{\pi}{\omega_p}}\frac{1}{\cos(\frac{\pi\omega_{p} L}{2})}\left[1-\frac{1}{\omega_p^2L^2}+\mathcal{O}\left(\frac{1}{L^4}\right)\right]\\&\times \int_{-\pi}^{\pi}\dd\tau'\,e^{\imath \omega_pL(\pi-\tau')}\mathcal{O}_-(\tau').
	\end{split}
\end{equation}

\hfill \break

\providecommand{\href}[2]{#2}\begingroup\raggedright\endgroup	

\begin{thebibliography}{10}
	\bibitem{Maldacena:1997re}
	J.~M. Maldacena, \emph{{The Large N limit of superconformal field theories and
			supergravity}}, \href{https://doi.org/10.1023/A:1026654312961}{\emph{Adv.
			Theor. Math. Phys.} {\bfseries 2} (1998) 231--252},
	[\href{https://arxiv.org/abs/hep-th/9711200}{{\ttfamily hep-th/9711200}}].
	
	\bibitem{Witten:1998qj}
	E.~Witten, \emph{{Anti-de Sitter space and holography}},
	\href{https://doi.org/10.4310/ATMP.1998.v2.n2.a2}{\emph{Adv. Theor. Math.
			Phys.} {\bfseries 2} (1998) 253--291},
	[\href{https://arxiv.org/abs/hep-th/9802150}{{\ttfamily hep-th/9802150}}].
	
	\bibitem{Gubser:1998bc}
	S.~S. Gubser, I.~R. Klebanov and A.~M. Polyakov, \emph{{Gauge theory
			correlators from noncritical string theory}},
	\href{https://doi.org/10.1016/S0370-2693(98)00377-3}{\emph{Phys. Lett. B}
		{\bfseries 428} (1998) 105--114},
	[\href{https://arxiv.org/abs/hep-th/9802109}{{\ttfamily hep-th/9802109}}].
	
	\bibitem{Freedman:1998tz}
	D.~Z.~Freedman, S.~D.~Mathur, A.~Matusis and L.~Rastelli,
	\emph{{Correlation functions in the CFT(d) / AdS(d+1) correspondence}},
	\href{https://doi.org/10.1016/S0550-3213(99)00053-X}{\emph{Nucl.~Phys.~B}
		{\bfseries 546} ~96~(1999)},
	[\href{https://arxiv.org/abs/hep-th/9804058}{{\ttfamily hep-th/9804058}}].
	
	\bibitem{Polchinski:1999ry}
	J.~Polchinski, \emph{{S matrices from AdS space-time}},
	[\href{https://arxiv.org/abs/hep-th/9901076}{{\ttfamily hep-th/9901076}}].
	
	\bibitem{Giddings:1999jq}
	S.~B. Giddings, \emph{{Flat space scattering and bulk locality in the AdS / CFT
			correspondence}},
	\href{https://doi.org/10.1103/PhysRevD.61.106008}{\emph{Phys. Rev. D}
		{\bfseries 61} (2000) 106008},
	[\href{https://arxiv.org/abs/hep-th/9907129}{{\ttfamily hep-th/9907129}}].
	
	\bibitem{Gary:2009ae}
	M.~Gary, S.~B. Giddings and J.~Penedones, \emph{{Local bulk S-matrix elements
			and CFT singularities}},
	\href{https://doi.org/10.1103/PhysRevD.80.085005}{\emph{Phys. Rev. D}
		{\bfseries 80} (2009) 085005},
	[\href{https://arxiv.org/abs/0903.4437}{{\ttfamily 0903.4437}}].
	
	\bibitem{Gary:2009mi}
	M.~Gary and S.~B. Giddings, \emph{{The Flat space S-matrix from the AdS/CFT
			correspondence?}},
	\href{https://doi.org/10.1103/PhysRevD.80.046008}{\emph{Phys. Rev. D}
		{\bfseries 80} (2009) 046008},
	[\href{https://arxiv.org/abs/0904.3544}{{\ttfamily 0904.3544}}].
	
	\bibitem{Heemskerk:2009pn}
	I.~Heemskerk, J.~Penedones, J.~Polchinski and J.~Sully, \emph{{Holography from
			Conformal Field Theory}},
	\href{https://doi.org/10.1088/1126-6708/2009/10/079}{\emph{JHEP} {\bfseries
			10} (2009) 079}, [\href{https://arxiv.org/abs/0907.0151}{{\ttfamily
			0907.0151}}].
	
	\bibitem{Fitzpatrick:2010zm}
	A.~L. Fitzpatrick, E.~Katz, D.~Poland and D.~Simmons-Duffin, \emph{{Effective
			Conformal Theory and the Flat-Space Limit of AdS}},
	\href{https://doi.org/10.1007/JHEP07(2011)023}{\emph{JHEP} {\bfseries 07}
		(2011) 023}, [\href{https://arxiv.org/abs/1007.2412}{{\ttfamily 1007.2412}}].
	
	\bibitem{Fitzpatrick:2011jn}
	A.~L. Fitzpatrick and J.~Kaplan, \emph{{Scattering States in AdS/CFT}},
	[\href{https://arxiv.org/abs/1104.2597}{{\ttfamily 1104.2597}}].
	
	\bibitem{Penedones:2010ue}
	J.~Penedones, \emph{{Writing CFT correlation functions as AdS scattering
			amplitudes}}, \href{https://doi.org/10.1007/JHEP03(2011)025}{\emph{JHEP}
		{\bfseries 03} (2011) 025},
	[\href{https://arxiv.org/abs/1011.1485}{{\ttfamily 1011.1485}}].
	
	\bibitem{Fitzpatrick:2011hu}
	A.~L. Fitzpatrick and J.~Kaplan, \emph{{Analyticity and the Holographic
			S-Matrix}}, \href{https://doi.org/10.1007/JHEP10(2012)127}{\emph{JHEP}
		{\bfseries 10} (2012) 127},
	[\href{https://arxiv.org/abs/1111.6972}{{\ttfamily 1111.6972}}].
	
	\bibitem{Okuda:2010ym}
	T.~Okuda and J.~Penedones, \emph{{String scattering in flat space and a scaling
			limit of Yang-Mills correlators}},
	\href{https://doi.org/10.1103/PhysRevD.83.086001}{\emph{Phys. Rev. D}
		{\bfseries 83} (2011) 086001},
	[\href{https://arxiv.org/abs/1002.2641}{{\ttfamily 1002.2641}}].
	
	\bibitem{Paulos:2016fap}
	M.~F. Paulos, J.~Penedones, J.~Toledo, B.~C. van Rees and P.~Vieira, \emph{{The
			S-matrix bootstrap. Part I: QFT in AdS}},
	\href{https://doi.org/10.1007/JHEP11(2017)133}{\emph{JHEP} {\bfseries 11}
		(2017) 133}, [\href{https://arxiv.org/abs/1607.06109}{{\ttfamily
			1607.06109}}].
	
	\bibitem{Hijano:2020szl}
	E. Hijano and D. Neuenfeld,
	\emph{{Soft photon theorems from CFT Ward identites in the flat limit of AdS/CFT}}, \href{https://doi.org/10.1007/JHEP11(2020)009}{\emph{JHEP} {\bfseries 11} (2020) 009},
	[\href{https://arxiv.org/abs/2005.03667}{{\ttfamily 2005.03667}}].
	
	\bibitem{Komatsu:2020sag}
	S.~Komatsu, M.~F. Paulos, B.~C. Van~Rees and X.~Zhao, \emph{{Landau diagrams in
			AdS and S-matrices from conformal correlators}},
	\href{https://doi.org/10.1007/JHEP11(2020)046}{\emph{JHEP} {\bfseries 11}
		(2020) 046}, [\href{https://arxiv.org/abs/2007.13745}{{\ttfamily
			2007.13745}}].
	
	\bibitem{Li:2021snj}
	Y.~Z.~Li,
	\emph{{Notes on flat-space limit of AdS/CFT}},
	\href{https://doi.org/10.1007/JHEP09(2021)027}{\emph{JHEP} {\bfseries 09}
		(2021) 027}, [\href{https://arxiv.org/abs/2106.04606}{{\ttfamily 2106.04606}}].
	
	\bibitem{Maldacena:2015iua}
	J.~Maldacena, D.~Simmons-Duffin and A.~Zhiboedov, \emph{{Looking for a bulk
			point}}, \href{https://doi.org/10.1007/JHEP01(2017)013}{\emph{JHEP}
		{\bfseries 01} (2017) 013},
	[\href{https://arxiv.org/abs/1509.03612}{{\ttfamily 1509.03612}}].
	
	\bibitem{Duary:2022pyv}
	S.~Duary, E.~Hijano and M.~Patra,
	\emph{{Towards an IR finite S-matrix in the flat limit of AdS/CFT}},
	[\href{https://arxiv.org/abs/2211.13711}{{\ttfamily 2211.13711}}].
	
	\bibitem{Duary:2022afn}
	S.~Duary,
	\emph{{AdS correction to the Faddeev-Kulish state: migrating from the flat peninsula}},
	\href{https://doi.org/10.1007/JHEP05(2023)079}{\emph{JHEP} {\bfseries 05}
		(2023) 079}, [\href{https://arxiv.org/abs/2212.09509 }{{\ttfamily 2212.09509}}].
	
	\bibitem{Li:2022tby}
	Y.~Z.~Li,
	\emph{{Flat-space structure of gluon and graviton in AdS}},
	[\href{https://arxiv.org/abs/2212.13195}{{\ttfamily 2212.13195}}].
	
	\bibitem{Hijano:2019qmi}
	E.~Hijano,
	\emph{{Flat space physics from AdS/CFT}},
	\href{https://doi.org/10.1007/JHEP07(2019)132}{\emph{JHEP} {\bfseries 07}
		(2019) 132}, [\href{https://arxiv.org/abs/1905.02729}{{\ttfamily 1905.02729}}].
	
	\bibitem{Li:2023azu}
	Y.~Z.~Li and J.~Mei,
	\emph{{Bootstrapping Witten diagrams via differential representation in Mellin space}},
	[\href{https://arxiv.org/abs/2304.12757}{{\ttfamily 2304.12757}}].
	
	\bibitem{Raju:2012zr}
	S.~Raju, \emph{{New Recursion Relations and a Flat Space Limit for AdS/CFT
			Correlators}}, \href{https://doi.org/10.1103/PhysRevD.85.126009}{\emph{Phys.
			Rev. D} {\bfseries 85} (2012) 126009},
	[\href{https://arxiv.org/abs/1201.6449}{{\ttfamily 1201.6449}}].
	
	\bibitem{Fitzpatrick:2011ia}
	A.~L.~Fitzpatrick, J.~Kaplan, J.~Penedones, S.~Raju and B.~C.~van~Rees,
	\emph{{A Natural Language for AdS/CFT Correlators}},
	\href{https://doi.org/10.1007/JHEP11(2011)095}{\emph{JHEP} {\bfseries 11}
		(2011) 095}, [\href{https://arxiv.org/abs/1107.1499}{{\ttfamily 1107.1499}}].
	
	\bibitem{Giombi:2017cqn}
	S.~Giombi, R.~Roiban, and A.~A. Tseytlin, {\it {Half-BPS Wilson loop and
			AdS$_2$/CFT$_1$}},  {\em Nucl. Phys.} {\bf B922} (2017) 499--527,
	[\href{http://arxiv.org/abs/1706.00756}{{\ttfamily 1706.00756}}].
	
	\bibitem{Beccaria:2019dws}
	M.~Beccaria, S.~Giombi and A.~A.~Tseytlin,
	\emph{{Correlators on non-supersymmetric Wilson line in $ \mathcal{N}=4 $ SYM and AdS$_{2}$/CFT$_{1}$}},
	\href{https://doi.org/10.1007/JHEP05(2019)122}{\emph{JHEP} {\bfseries 05}
		(2019) 122}, [\href{https://arxiv.org/abs/1903.04365}{{\ttfamily 1903.04365}}].
	
	\bibitem{Beccaria:2019stp}
	M.~Beccaria and A.~A.~Tseytlin,
	\emph{{On boundary correlators in Liouville theory on AdS$_{2}$}},
	\href{https://doi.org/10.1007/JHEP07(2019)008}{\emph{JHEP} {\bfseries 07} (2019) 008}, [\href{https://arxiv.org/abs/1903.04365}{{\ttfamily 1904.12753}}].
	
	\bibitem{Beccaria:2019ibr}
	M.~Beccaria and G.~Landolfi,  \emph{{Toda theory in AdS$_{2}$ and $\mathcal
			WA_{n}$-algebra structure of boundary correlators}},
	\href{https://doi.org/10.1007/JHEP10(2019)003}{\emph{JHEP} {\bfseries 07} (2019) 008}, [\href{https://arxiv.org/abs/1906.06485}{{\ttfamily 1906.06485}}].
	
	\bibitem{Beccaria:2019mev}
	M.~Beccaria, H.~Jiang, and A.~A. Tseytlin, \emph{{Non-abelian Toda theory on
			AdS$_2$ and AdS$_2$/CFT$_2^{1/2}$ duality}},
	\href{http://dx.doi.org/10.1007/s13130-019-11219-y}{{\em JHEP} {\bfseries 09}
		(2019) 036}, [\href{http://arxiv.org/abs/1907.01357}{{\ttfamily 1907.01357}}].
	
	\bibitem{Beccaria:2019dju}
	M.~Beccaria, H.~Jiang, and A.~A. Tseytlin, \emph{{Supersymmetric Liouville theory
			in AdS$_{2}$ and AdS/CFT}},
	\href{http://dx.doi.org/10.1007/JHEP11(2019)051}{{\em JHEP} {\bfseries 11}
		(2019) 051}, [\href{http://arxiv.org/abs/1909.10255}{{\ttfamily 1909.10255}}].
	
	\bibitem{Beccaria:2020qtk}
	M.~Beccaria, H.~Jiang, and A.~A. Tseytlin, \emph{{Boundary correlators in WZW model on AdS$_{2}$}}, \href{http://dx.doi.org/10.1007/JHEP05(2020)099}{{\em JHEP}
		{\bfseries 05} (2020) 099}, [\href{http://arxiv.org/abs/2001.11269}{{\ttfamily
			2001.11269}}].
	
	\bibitem{Gadde:2022ghy}
	A.~Gadde and T.~Sharma,
	\emph{{A scattering amplitude for massive particles in AdS}}, \href{http://dx.doi.org/10.1007/JHEP09(2022)157}{{\em JHEP}
		{\bfseries 09} (2022) 157}, [\href{http://arxiv.org/abs/2204.06462 }{{\ttfamily
			2204.06462}}].
	
	\bibitem{Banks:1998dd}
	T.~Banks, M.~R.~Douglas, G.~T.~Horowitz and E.~J.~Martinec,
	\emph{{AdS dynamics from conformal field theory}},
	[\href{https://arxiv.org/abs/hep-th/9808016}{{\ttfamily hep-th/9808016}}].
	
	\bibitem{Hamilton:2005ju}
	A.~Hamilton, D.~N. Kabat, G.~Lifschytz and D.~A. Lowe, \emph{{Local bulk
			operators in AdS/CFT: A Boundary view of horizons and locality}},
	\href{https://doi.org/10.1103/PhysRevD.73.086003}{\emph{Phys. Rev. D}
		{\bfseries 73} (2006) 086003},
	[\href{https://arxiv.org/abs/hep-th/0506118}{{\ttfamily hep-th/0506118}}].
	
	\bibitem{Hamilton:2006az}
	A.~Hamilton, D.~N. Kabat, G.~Lifschytz and D.~A. Lowe, \emph{{Holographic
			representation of local bulk operators}},
	\href{https://doi.org/10.1103/PhysRevD.74.066009}{\emph{Phys. Rev. D}
		{\bfseries 74} (2006) 066009},
	[\href{https://arxiv.org/abs/hep-th/0606141}{{\ttfamily hep-th/0606141}}].
	
	\bibitem{Cordova:2022pbl}
	L.~C\'ordova, Y.~He and M.~F.~Paulos,
	\emph{{From conformal correlators to analytic S-matrices: CFT$_{1}$/QFT$_{2}$}},
	\href{https://doi.org/10.1007/JHEP08(2022)186}{\emph{JHEP}
		{\bfseries 08} (2022) 186},
	[\href{https://arxiv.org/abs/hep-th/2203.10840}{{\ttfamily 2203.10840}}].
	
	\bibitem{Duary:2022onm}
	S.~Duary,
	\emph{{Celestial amplitude for 2d theory}},
	\href{https://doi.org/10.1007/JHEP12(2022)060}{\emph{JHEP}
		{\bfseries 12} (2022) 060},
	[\href{https://arxiv.org/abs/hep-th/2209.02776}{{\ttfamily 2209.02776}}].
	
	\bibitem{Kapec:2022xjw}
	D.~Kapec and A.~Tropper,
	\emph{{Integrable field theories and their CCFT duals}},
	\href{https://doi.org/10.1007/JHEP02(2023)128}{\emph{JHEP}
		{\bfseries 02} (2023) 128},
	[\href{https://arxiv.org/abs/hep-th/2210.16861}{{\ttfamily 2210.16861}}].
	
	\bibitem{Zamolodchikov:1978xm}
	A.~B.~Zamolodchikov and A.~B.~Zamolodchikov,
	\emph{{Factorized S matrices in two-dimensions as the exact solutions of certain relativistic quantum field modelss}}, \href{https://doi.org/10.1016/0003-4916(79)90391-9}{Annals Phys. \textbf{120}, 253 (1979)}.
	
	\bibitem{Dorey:1996gd}
	P.~Dorey,
	\emph{{Exact S matrices}},
	[\href{https://arxiv.org/abs/hep-th/9810026}{{\ttfamily hep-th/9810026}}].
	
	\bibitem{Parke:1980ki}
	S.~J.~Parke,
	\emph{{Absence of Particle Production and Factorization of the $S$ Matrix in (1+1)-dimensional Models}}, \href{https://doi.org/10.1016/0550-3213(80)90196-0}{Nucl. Phys. B \textbf{174}, 166 (1980)}. 
	
	\bibitem{Paulos:2016but}
	M.~F.~Paulos, J.~Penedones, J.~Toledo, B.~C.~van Rees and P.~Vieira,
	\emph{{The S-matrix bootstrap II: two dimensional amplitudes}},
	\href{http://dx.doi.org/10.1007/JHEP11(2017)143}{{\em JHEP} {\bfseries 11}
		(2017) 143}, [\href{http://arxiv.org/abs/1607.06110}{{\ttfamily 1607.06110}}].
	
	\bibitem{Rosenhaus:2019utc}
	V.~Rosenhaus and M.~Smolkin,
	\emph{{Integrability and renormalization under $T \bar T$}},
	\href{https://doi.org/10.1103/PhysRevD.102.065009}{\emph{Phys. Rev. D}
		{\bfseries 102} (2020) no.6, 065009},
	[\href{https://arxiv.org/abs/1909.02640}{{\ttfamily 1909.02640}}].
	
	\bibitem{Antunes:2021abs}
	A.~Antunes, M.~S.~Costa, J.~Penedones, A.~Salgarkar and B.~C.~van Rees,
	\emph{{Towards bootstrapping RG flows: sine-Gordon in AdS}},
	\href{http://dx.doi.org/10.1007/JHEP12(2021)094}{{\em JHEP} {\bfseries 12}
		(2021) 094}, [\href{http://arxiv.org/abs/2109.13261}{{\ttfamily 2109.13261}}].
	
	\bibitem{Ouyang:2019xdd}
	H.~Ouyang,
	\emph{{Holographic four-point functions in Toda field theories in
			AdS$_{2}$}},
	\href{http://dx.doi.org/10.1007/JHEP04(2019)159}{{\em JHEP} {\bfseries 04}
		(2019) 159}, [\href{http://arxiv.org/abs/1607.06110}{{\ttfamily 1902.10536}}].
	
	\bibitem{Mazac:2016qev}
	D.~Mazac,
	\emph{{Analytic bounds and emergence of AdS$_{2}$ physics from the
			conformal bootstrap}},
	\href{http://dx.doi.org/10.1007/JHEP04(2017)146}{{\em JHEP} {\bfseries 04}
		(2017) 146}, [\href{http://arxiv.org/abs/1611.10060}{{\ttfamily 1611.10060}}].
	
	\bibitem{Ferrero:2019luz}
	P.~Ferrero, K.~Ghosh, A.~Sinha and A.~Zahed,
	\emph{{Crossing symmetry, transcendentality and the Regge behaviour of 1d
			CFTs}},
	\href{http://dx.doi.org/10.1007/JHEP07(2020)170}{{\em JHEP} {\bfseries 07}
		(2020) 170}, [\href{http://arxiv.org/abs/1911.12388}{{\ttfamily 1911.12388}}].
	
	\bibitem{DiPietro:2017vsp}
	L.~Di~Pietro and E.~Stamou,
	\emph{{Operator mixing in the $\boldsymbol{\epsilon}$-expansion: Scheme and
			evanescent-operator independence}},
	\href{https://doi.org/10.1103/PhysRevD.97.065007}{\emph{Phys. Rev. D}
		{\bfseries 97} (2018), 065007},
	[\href{https://arxiv.org/abs/1708.03739}{{\ttfamily 1708.03739}}].
	
	\bibitem{Maldacena:1998uz}
	J.~M.~Maldacena, J.~Michelson and A.~Strominger,
	\emph{{Anti-de Sitter fragmentation}},
	\href{http://dx.doi.org/10.1007/JHEP07(2020)170}{{\em JHEP} ~9902,~011~(1999)},
	[\href{https://arxiv.org/abs/hep-th/9812073}{{\ttfamily hep-th/9812073}}].
	
	\bibitem{Gross:2017vhb}
	D.~J.~Gross and V.~Rosenhaus,
	\emph{{A line of CFTs: from generalized free fields to SYK}},
	\href{http://dx.doi.org/10.1007/JHEP07(2017)086}{{\em JHEP} {\bfseries 07}
		(2017) 086}, [\href{http://arxiv.org/abs/1706.07015}{{\ttfamily 1706.07015}}].
	
	
	\bibitem{Maldacena:2016hyu}
	J.~Maldacena and D.~Stanford,
	\emph{{Remarks on the Sachdev-Ye-Kitaev model}},
	\href{https://doi.org/10.1103/PhysRevD.94.106002}{\emph{Phys. Rev. D}
		{\bfseries 94} (2016) 106002}, [\href{https://arxiv.org/abs/1604.07818}{{\ttfamily 1604.07818}}].
	
\end{thebibliography}
\end{document}